\documentclass[preprint]{aastex}
\newcommand\blfootnote[1]{%
  \begingroup
  \renewcommand\thefootnote{}\footnote{#1}%
  \addtocounter{footnote}{-1}%
  \endgroup
}
\usepackage{amssymb,amsmath,natbib,graphicx,bm}
\usepackage[T1]{fontenc}

\newcommand{\Oe}{OE}%{outer east}
\newcommand{\Ow}{OW}%{outer west}
\newcommand{\Iw}{IW}%{innerwest}
\newcommand{\Ie}{IE}%{inner east}
\newcommand{\C}{C}%{central}
\newcommand{\hii}{H{\rmfamily\scshape{ii}}}
\newcommand{\kms}{\mbox{km~s$^{-1}$}}
\newcommand{\Msun}{\mbox{\,$M_{\odot}$}}
\newcommand{\Lsun}{\mbox{$L_{\odot}$}}
\shortauthors{Guzm{\'a}n et al.}
\shorttitle{Protostellar Jet from a HC \hii R}

\slugcomment{\today}

\begin{document}
%%%%%%%%%%%%%%%%%%%%%%%%%%%%%%%%%%%%%%%%%%%%%%%%%%%%%%%%%%%%%%%%%%%%%%%%%
% TITLE and AUTHORS
%%%%%%%%%%%%%%%%%%%%%%%%%%%%%%%%%%%%%%%%%%%%%%%%%%%%%%%%%%%%%%%%%%%%%%%%%

\title{A Protostellar Jet Emanating from a Hypercompact \hii\ Region}
\author{Andr\'es E. Guzm\'an\altaffilmark{1}, Guido Garay\altaffilmark{1},
  Luis F.\ Rodr\'iguez\altaffilmark{2}, Yanett Contreras\altaffilmark{3,4}, Catherine
  Dougados\altaffilmark{1}, Sylvie Cabrit\altaffilmark{5}}

\altaffiltext{1}{Departamento de Astronom\'{\i}a, Universidad de Chile,
  Camino el Observatorio 1515, Las Condes, Santiago, Chile}
\altaffiltext{2}{Instituto de Radioastronom{\'\i}a y Astrof{\'\i}sica (UNAM), Morelia 58089, M\'exico}
\altaffiltext{3}{CSIRO Astronomy and Space Science, P.O. Box 76, Epping
  1710 NSW, Australia}
\altaffiltext{4}{Leiden Observatory, Leiden University, PO Box 9513, NL-2300 RA Leiden, the Netherlands}
\altaffiltext{5}{LERMA, Observatoire de Paris, UMR 8112 du CNRS, ENS, UPMC, UCP, 61 Av.\ de l'Observatoire, F-75014 Paris, France}

%%%%%%%%%%%%%%%%%%%%%%%%%%%%%%%%%%%%%%%%%%%%%%%%%%%%%%%%%%%%%%%%%%%%%%%%%
% Abstract
%%%%%%%%%%%%%%%%%%%%%%%%%%%%%%%%%%%%%%%%%%%%%%%%%%%%%%%%%%%%%%%%%%%%%%%%%
\begin{abstract}
We present radio continuum observations of the high-mass young stellar
object (HMYSO)  G345.4938+01.4677   made  using  the  \emph{Australia
  Telescope Compact  Array} (ATCA) at  5, 9,  17, and  19 GHz. These observations 
 provide  definite evidence that the outer and inner pairs of radio lobes consist
of shock  ionized material being  excited by an  underlying collimated
and fast protostellar jet emanating from a hypercompact \hii\ region.
By comparing with images  taken 6 yr earlier at 5 and  9 GHz using the
same telescope,  we assess  the proper motions  of the  radio sources.
The outer West and East lobes  exhibit proper motions of $64\pm12$ and
$48\pm13$  milliarcsec yr$^{-1}$,  indicating velocities  projected in
the plane  of the sky and  receding from G345.4938+01.4677 of  520 and
390 \kms, respectively.
The  internal   radio  lobes   also  display  proper   motion  signals
consistently receding from the HMYSO, with magnitudes of $17\pm11$ and
$35\pm10$ milliarcsec  yr$^{-1}$ for  the inner  West and  East lobes,
respectively.
The morphology  of the outer West  lobe is that  of a detached
bow shock.  At 17 and 19 GHz, the outer East lobe displays an arcuate
morphology also suggesting a bow shock.
%
%We propose that the observed bending of the jet could be explained as
%due to the displacement of the HMYSO through the molecular clump.
%
%
These results show that disk  accretion and jet acceleration --- possibly
occurring in a very similar  way compared with low-mass protostars ---  is
taking  place in  G345.4938+01.4677 despite  the presence  of ionizing
radiation and the associated hypercompact \hii\ region.
\end{abstract}


\section{INTRODUCTION}

In spite of the preponderant influence of high-mass stars ($M_\star>8$ \Msun)
on the evolution of galaxies, with them being the progenitors of core collapse
supernovae and the main sources of chemical enrichment, turbulence and
mixing in the interstellar medium, the details of their formation are not
well understood \citep{Zinnecker2007ARA&A,Tan2014PrPl}.  The formation of
high-mass stars is more difficult to study than that of low-mass stars due
to their relative scarcity, larger distances, shorter formation timescales ($\le 10^6$ yr),
and large extinction toward their birth places ($A_V>10$).

The currently accepted evolutionary path of high-mass stars begins inside
dense and massive molecular cores, where high-mass young stellar objects
(HMYSOs) accrete at rates between $10^{-5}$ to $10^{-3}$ \Msun\ yr$^{-1}$
\citep{Tan2014PrPl}.  The young stars finish their Kelvin-Helmholtz
contraction very rapidly and reach the main sequence
\citep{Keto2006ApJ,Norberg2000AA}. At this point, the star radiates extreme
ultraviolet (UV) photons that ionize its surroundings, producing a
hypercompact (HC) \hii\ region.  HC \hii\ regions are observationally
characterized by sizes $\la0.03$ pc, densities $n_e>10^6$ cm$^{-3}$,
emission measures $>10^8$ pc cm$^{-6}$, and hydrogen recombination line
(HRL) widths $\gtrsim50$ \kms\ \citep{Kurtz2000RMxAACS,Hoare2007PrPlV}.
Eventually, the accretion flow will stop and the UV radiation and winds
from the recently formed star will disperse the rest of the envelope
material, leaving the O-type star surrounded by a classical \hii\ region
\citep{Zapata2010ApJ}.

Recent observational evidence gathers in favor of accretion as the
preferred high-mass star formation mechanism despite the radiation
pressure, the thermal pressure of the heated protostellar material, and the
ionized gas pressure.  These processes in combination could reverse,
theoretically, the accretion flows \citep{Larson1971AA,Wolfire1987ApJ}.
High accretion rates have been invoked in high-mass star formation in order
to overcome the radiation pressure and to choke the development of an
\hii\ region \citep{Garay1999PASP}.  However, what appears to be crucial in
the solution of the theoretical problems described above is the
non-spherical character of accretion.  
Disk accretion, in particular, is an effective way to circumvent the
radiation and ionized gas pressure, allowing the accreting material to
reach the young high-mass stars much more easily by flowing inwards
mainly through the plane perpendicular to the angular momentum vector
of the system \citep{Nakano1989ApJ,Kuiper2011ApJ}.  Confirming this
theoretical picture entails finding observational evidence of disk
accretion toward HMYSOs.  Currently, most of the evidence is indirect
and consists of contracting motions of molecular clumps, poorly
collimated molecular outflows, and rotationally flattened molecular
structures (see \citealp{Guzman2014ApJ} and references therein).

For O-type stars we expect that a fraction $\gtrsim50\%$ of
the final stellar mass to be accreted after the Kelvin-Helmholtz
contraction and after the star starts to produce ionizing radiation
\citep{Hosokawa2009ApJ,Zhang2014ApJ}.  If material during this stage is
accreted through a disk, then it is pertinent to look for evidence of disk
accretion toward HC \hii\ regions and to settle the question as to whether
or not accretion onto the HMYSO is maintained after stellar contraction and
UV photon injection. Note that accretion through a disk may choke the
ionized region only near the disk plane, while allowing the \hii\ region
development in the polar regions \citep[][]{Keto2007ApJ}.  HC \hii\ regions
indeed exhibit indirect evidence pointing in this direction, such as very
high column densities ($N_{H_2}>10^{23}$ cm$^{-2}$), high incidence
($>70\%$) of infalling motions of the surrounding molecular gas
\citep{Klaassen2012AA}, molecular outflows, rotationally flattened
molecular structures, and time-varying radio
fluxes \citep{Galvan-Madrid2008ApJ}.

Currently, the most direct way to assert that a high-mass star is
accreting from a disk is by finding highly collimated jets (aperture
$\le5\arcdeg$) with velocities comparable with the escape velocity of
the central object \citep{Livio2009pjc}. While low-velocity and poorly
collimated outflows are produced by a variety of processes like
magnetic braking \citep{Hennebelle2011AA}, ionization feedback
\citep{Peters2012ApJ}, and even mergers \citep{Bally2005AJ}, highly
collimated and fast outflows seem to be more reliable signposts of
disk accretion \citep{Seifried2012MNRAS}. Observationally, collimated
jets \citep[e.g.,][]{Purser2016arXiv,Caratti-o-Garatti2015AA} and
keplerian-like disks
\citep[e.g.,][]{Johnston2015ApJ,Sanchez-Monge2013AA,Kraus2010Nat} have been
observed associated with luminous HMYSOs, supporting the disk
accretion scenario for high-mass stars.

In this work we present new observations of the multiple radio source
detected toward the HMYSO G345.4938+01.4677 (G345.49+1.47
hereafter).\footnote{We use the nomenclature of the  RMS survey
  (http://www.ast.leeds.ac.uk/RMS, \citealp{Lumsden2013ApJ}) 
based on the \emph{Midcourse Space Experiment} coordinates.} 
With a bolometric luminosity of
$\sim32,000\,\Lsun$\ \citep[assuming a distance of 1.7
  kpc,][]{Lopez2011AA}, G345.49+1.47 is the dominating central HMYSO
of the $\sim1000\Msun$ molecular clump associated with IRAS
16562$-$3959. The bolometric luminosity of the IRAS source is
$5$--$7\times10^4\Lsun$
\citep{Lopez2011AA}.\footnote{Note that the RMS survey adopts a
  distance of 2.4 kpc and a luminosity of $1.5\times10^5$ \Lsun\ for this HMYSO.}
Five roughly aligned radio continuum sources were detected toward
G345.49+1.47 by \citet{Guzman2010ApJ}.  Four of these flank in an
approximately symmetric way a bright central radio source located at
the position of the HMYSO.  In addition, infrared (IR) 2MASS and
GLIMPSE images show emission extending toward the east and along the
direction of the string of radio sources.  \citet{Guzman2010ApJ}
interpreted the emission from the central radio source as arising from
an ionized protostellar jet and the rest of the sources as
shock-ionized lobes.  The IR appearance G345.49+1.47 can be understood
as the illuminated inner walls of an outflow cavity which contains the
jet. This emission is enhanced toward the eastern side of the jet
which corresponds --- as inferred from  observations of an associated molecular bipolar
outflow \citep{Guzman2011ApJ} --- to the blue-shifted side
of the jet.  Similar interpretations have been proposed for other
well-know HMYSOs associated with radio jets such as G343.1262$-$00.0620  \citep[also IRAS
  16547$-$4247,][]{Garay2003ApJ,Rodriguez2008AJ}.

Contrasting these interpretations, HRL observations of G345.49+1.47 \citep{Guzman2014ApJ}
demonstrated that the central source is not a protostellar ionized jet but
a photoionized HC \hii\ region.  The question arises as to whether or not
the radio lobes are being excited by an underlying (and still undetected) jet, which would be 
part of a purported disk-jet system within this HC \hii\ region.
%The present
%work seeks to provide  evidence for an accreting disk
%coexisting with photoionized gas and UV radiation. 
Evidence in this direction will support the adequacy of the disk-jet star formation
mechanism in the high-mass case, even after the onset of ionizing
radiation.
Section \ref{sec-obs} of this work describes the new observations performed
toward G345.49+1.47. Section \ref{sec-res} presents the results, and
Section \ref{sec-dis} discusses the interpretation and jet characteristics.
We summarize our conclusions in Section \ref{sec-sum}.

{\section{OBSERVATIONS}\label{sec-obs}}

We observed G345.49+1.47 using the \emph{Australia Telescope Compact
  Array}\footnote{The Australia Telescope Compact Array is part of the
  Australia Telescope National Facility which is funded by the Commonwealth
  of Australia for operation as a National Facility managed by CSIRO.}
(ATCA) in October of 2014 and in March and May of 2015 under the project ID number 
C3006. We observed three
runs of 12 hrs, one using the 1.5 km configuration and the other two the
6.0 km configuration. The phase center of the array was toward
R.A.$=16^{\rm h}$59$^{\rm m}$41\fs61, $\text{decl.}=-40\arcdeg$03\arcmin43\farcs4 (J2000).
We split evenly the time among two frequency settings using the
\emph{Compact Array Broadband Backend} \citep{CABB} without zooms.
The first frequency setting covers simultaneously the 4.0--6.0 GHz and the
8.0--10.0 GHz spectral windows using 2048 channels of 1 MHz width.  The
second frequency setting covers simultaneously the 16.0-18.0 GHz and the
18.0--20.0 GHz spectral windows, also using 2048 channels of 1 MHz width. 
Hereafter, we refer to each of these four
frequency intervals as 5, 9, 17 and 19 GHz, respectively.

\begin{deluxetable}{lcccccccc}
\tablewidth{0pc} \tablecolumns{9} \tabletypesize{\footnotesize}
\tablecaption{Observational Parameters\label{tab-noise}}
\tablehead{
\colhead{Epoch} &\multicolumn{4}{c}{Synthesized Beam (BMAJ $\times$ BMIN, P.A.)}&\multicolumn{4}{c}{Noise }\\
\colhead{}&\multicolumn{4}{c}{($\arcsec\times\arcsec, \arcdeg$)}& \multicolumn{4}{c}{($\mu$Jy beam$^{-1}$)}\\
\colhead{}&\colhead{5 GHz} & \colhead{9 GHz}& \colhead{17 GHz} & \colhead{19 GHz}&
\colhead{5 GHz} & \colhead{9 GHz}& \colhead{17 GHz} & \colhead{19 GHz}
}
\startdata
2008/09 &$2.74\times1.74, 2$&$1.59\times1.05, 3$&\nodata&\nodata& 91 & 75  & \nodata & \nodata \\
2014/15 &$2.49\times1.36, 3$&$1.59\times0.89, 3$&$0.83\times0.44, 6$&$0.78\times0.39, -3$& 18 & 17&30 &27\\
\enddata
\end{deluxetable}

We calibrated the data using the \emph{Miriad} software
\citep{Sault1995ASPC}. The flux scale was determined by comparing with the
calibrator PKS 1934$-$638. We corrected for atmospheric and antenna based
gain variations by observing the gain calibrator PKS 1729$-$37 regularly
every 8 min. The fluxes measured toward PKS 1729$-$37 were 1.19, 1.27, 1.31,
and 1.29 Jy at 5, 9, 17, and 19 GHz, respectively, which are within the
observed flux
variations.\footnote{http://www.narrabri.atnf.csiro.au/calibrators/}
Deconvolution and imaging was performed using the \emph{Common Astronomy
  Software Applications} \citep[CASA,][]{Reid2010AAS}.

We also re-reduced archival data\footnote{This paper includes archived
  data obtained through the Australia Telescope Online Archive
  (http://atoa.atnf.csiro.au).}  taken between October 2008 and
February 2009 toward G345.49+1.47 and originally published by
\citet{Guzman2010ApJ}.  The source PKS 1740$-$517 was used as gain
calibrator.  Since the radio emission toward G345.49+1.47 is dominated
by a central, compact source of $\sim13$ mJy at 9 GHz with an spectral
index of $\sim1$ (see Section \ref{sec-res}), we were able to apply
phase self-calibration. This new reduction and imaging of the 9 GHz
data reaches a similar noise level as the one presented in
\citet{Guzman2010ApJ} but with less artifacts. Particularly, we do not
recover the three aligned sources mentioned by \citet{Guzman2010ApJ}
located 4\arcsec\ north of their outer east lobe. Therefore, and
also because these sources were aligned along one of the ridges of 
the dirty beam pattern, we conclude that they
are likely spurious.

Table \ref{tab-noise} gives the synthesized beams and the noise (measured
as the rms deviations in areas devoid of sources) of the images taken at 5,
9, 17, and 19 GHz. The first and second rows of Table \ref{tab-noise} list
the parameters of the data taken between 2008--2009 and 2014--2015,
respectively. We refer hereafter 
to these two epochs as 2008/09 and 2014/15. Note
that no data were taken at 17 and 19 GHz  in the 2008/09 epoch. The
sensitivity difference between the two epochs is due primarily to the
backend bandwidth, being 128 MHz in the 2008/09 observations versus
$\sim2$ GHz in the 2014/15 epoch. Fully calibrated and reduced images are
available through the \emph{Dataverse}.\footnote{http://dx.doi.org/10.7910/DVN/EMZPWU}

%\afterpage{\clearpage}%
{\section{RESULTS}\label{sec-res}}

\begin{figure}
\centering\includegraphics[width=0.95\textwidth]{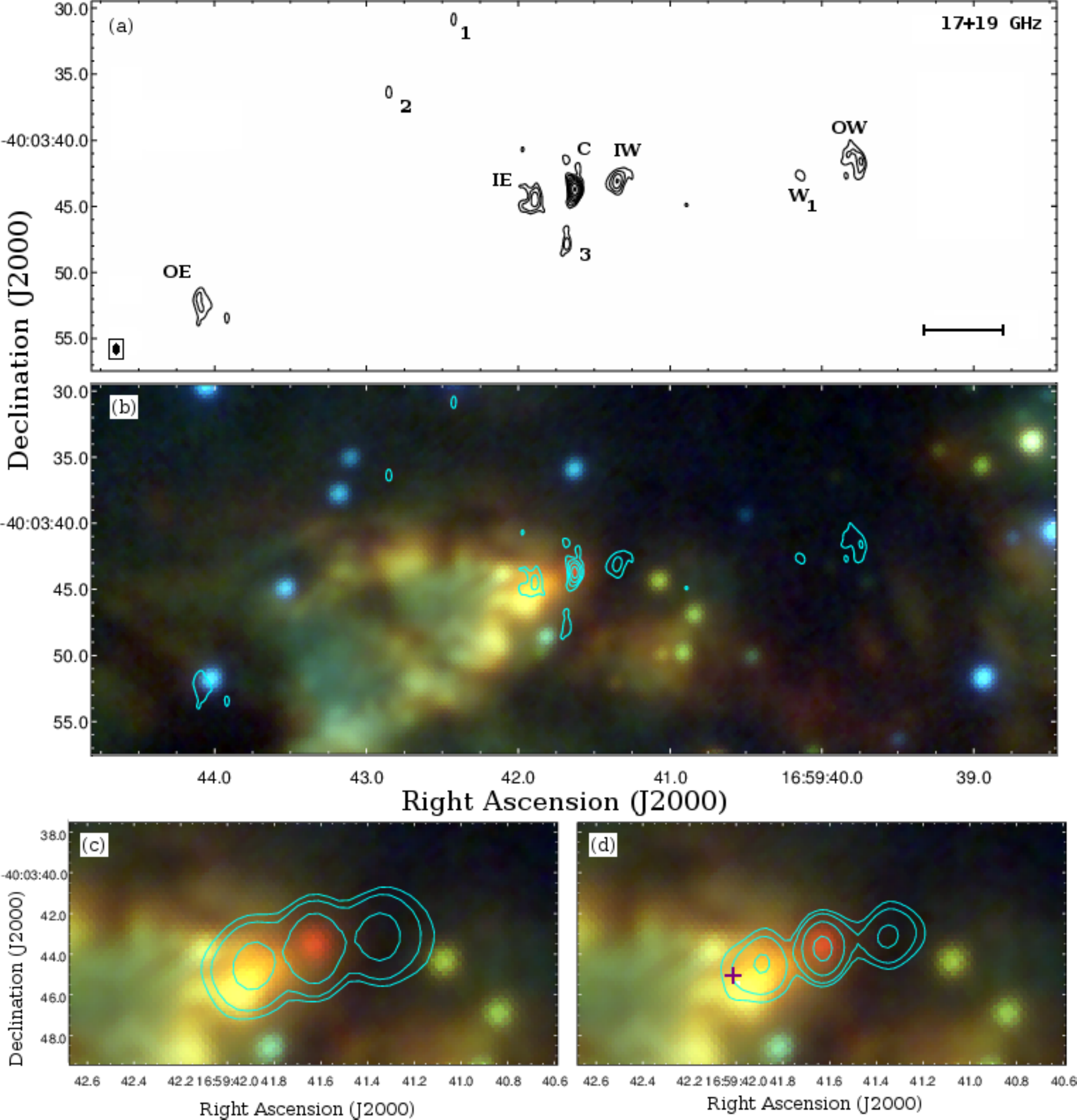}
\caption{\small\baselineskip=0.64\baselineskip Panel (a): contour map of
  the radio continuum emission observed at 17 and 19 GHz combined. Contour
  levels: $7\sigma\times~2^i$ for $i=1\ldots8$ and $\sigma=20\mu$Jy. Labels
  are attached to each of the sources (Tables \ref{tab-pos} and
  \ref{tab-flux}). The bar located in the bottom-right corner of the map
  measures 0.05 pc at 1.7 kpc. Panel (b): three-color image of G345.49+1.47
  made using J (blue), H (green), and K$_S$ (red) data from the VVV survey.
  Cyan contours are as in panel (a) but drawn at every second level
  starting from the first. Panels (c) and (d) show a zoom-in into the
  region adjacent to G345.49+1.47 with superimposed cyan contours of the 5
  and 9 GHz data, respectively. Contour levels for panels (c) and (d) are
  given by $(3+4^i)\times50~\mu$Jy with $i=0\ldots2$ and $i=0\dots3$,
  respectively. The purple cross in panel (d) indicates the position of
  2MASS 16594202$-$4003450. \label{fig-pres}}
\end{figure}

The upper panel of Figure \ref{fig-pres} shows a contour map of the 17
and 19 GHz data combined taken during the 2014/15 epoch with labels
for each of the detected sources.  Clearly detected are the five
sources reported by \citet{Guzman2010ApJ}, named outer east, inner
east, central, inner west, and outer west and marked as OE, IE, C, IW,
and OW, respectively. Roughly aligned with these sources and
$\sim5\arcsec$ east of OW we detect an additional knot denoted W$_1$.
In addition, we detect other three sources in the field numbered from
1 to 3 in decreasing declination order.  These sources were below the
4-$\sigma$ level in the 9 GHz images of \citet{Guzman2010ApJ} and
therefore they were not reported.  The W$_1$ knot is detected in the
2008/09 9 GHz image at $\sim5$-$\sigma$ but, due to the lower
fidelity of the  images presented in \citet{Guzman2010ApJ}, it was not reported 
as a separate source.

Figure \ref{fig-pres} also shows (middle panel) a three-color near-IR image of G345.49+1.47 made using J, H,
and K$_S$ band observations  obtained from the \emph{Vista
  Variables in the Via-Lactea}
survey\footnote{http://vvvsurvey.drupalgardens.com/}
\citep[VVV,][]{Minniti2010NewA} public data.  
The three-color image shows the
scattered light from the inner walls of the outflow cavity associated with
the blue-shifted side of the bipolar outflow 
%and to a blue-shifted OH 1720 MHz maser emission 
\citep{Guzman2011ApJ}.
The three-color image also shows a near-IR source detected in K$_S$ band whose
position ($\text{R.A.}=16^{\rm h}59^{\rm m}41\fs645$,
$\text{decl.}=-40^{\rm d}03^{\rm m}43\farcs64$) is coincident with the
central radio source C and has an estimated apparent magnitude
of K${_S}\approx12.4$. 
This near-IR source --- better displayed in panels
(c) and (d) ---  is likely the near-IR counterpart of the embedded HMYSO
G345.49+1.47, associated also with the  mid-IR source \mbox{G345.4938+01.4677-Src\#2} detected by \citet{Mottram2007AA}. 
Note that the nearest 2MASS source (16594202$-$4003450) is located toward the  
bright tip of the outflow cavity rather than toward source C.
%
%% %
%% According to the
%% stellar parameters given by \citet{Mottram2011ApJL} and assuming that the
%% HMYSO mass is between 15 and 20 \Msun, we derive an intrinsic K$_S$
%% magnitude for G345.49+1.47 ranging between $9.8$ and $9.4$. 
%% %
%% Interpreting
%% the K$_S$ emission as absorbed photospheric light, we derive a visual extinction
%% $(A_V)$ toward the HMYSO ranging between $23$ and $27$, assuming
%% $A_{K_S}/A_V=0.112$ \citep{Rieke1985ApJ}. 
%% %
%% Note that these values are likely
%% lower limits, since a fraction of the emission in the K$_S$ band could be
%% instead being re-emitted by circumstellar material.

%% pdl> p (2.5*log10(251070-127491)+11.244)
%% 23.97
%% pdl> p (10.804+2.5*log10(687618 - 417886))
%% 24.3813311814735
%pdl> p (2.5*log10(917787-419291)+9.951)
%24.1951541945585
%pdl>  p (-2.5*log10(190686-139906)+(23.97* 24.38133118147*24.19515419)**(1/3))
% 12.4173449254562

\subsection{Flux Density and Morphology of the Radio Emission}

The five radio sources detected toward G345.49+1.47 during the 2008/09
epoch are detected unambiguously in the 2014/15 images.  We were also able to
detect at a level of 7-$\sigma$ or better four additional sources due to
the improved sensitivity of the 2014/15 observations: those marked in
Figure \ref{fig-pres} with numbers 1, 2, and 3, and knot W$_1$, situated at
$17\farcs1$ from source C in the $\text{P.A.}=-87\arcdeg$ direction. Knot
W$_1$ is located between sources IW and OW and is approximately aligned
with them.

Tables \ref{tab-pos} and \ref{tab-flux} list the positions and primary beam
corrected flux densities,\footnote{\raggedright ATCA primary beam responses
  obtained from http://www.narrabri.atnf.csiro.au/observing/users\_guide}
respectively, of all the detected sources.  We used
2D-Gaussian fittings to calculate the positions and the uncertainties of
sources 1, 2, 3, W$_1$, C, and of the inner lobes using the \texttt{imfit}
task within CASA. The uncertainty in the position of a Gaussian source is
given approximately by \citep{Condon1997PASP}
\begin{equation}\label{eq-ung}
\eta\sigma_G\left(\sqrt{N_{\rm eff}}~{\rm SNR_p}\right)^{-1}~~, 
\end{equation}
where $\sigma_G$ is the position standard deviation of the Gaussian
($\sigma_G=\theta_{\rm FWHM}/2\sqrt{2\ln2}$), $N_{\rm eff}$ is the quotient
between the source and beam solid angles, ${\rm SNR_p}$ is the peak
intensity of the source divided by the rms deviation of the image, and
$1<\eta<2\sqrt{2}$ is a correction factor that accounts for noise
correlation on scales comparable with the beam size. For unresolved sources,
$\eta\sim1$.
\begin{deluxetable}{lccccc}
\rotate
\tablewidth{0pt} \tablecolumns{6} \tabletypesize{\scriptsize}
\tablecaption{Position of the Radio Sources\label{tab-pos}}
\tablehead{
\colhead{Source} & \colhead{Epoch} & \multicolumn{4}{c}{Right Ascension and Declination (J2000)}\\
\colhead{ } & \colhead{ } & \multicolumn{4}{c}{16:59:\ldots,~$-$40:03:\ldots}\\
\colhead{} & \colhead{} & \colhead{5 GHz} & \colhead{9 GHz}& \colhead{17 GHz} & \colhead{19 GHz}
}
\startdata
C       & 2008/09 & $41\fs623\pm0\fs001,~43\farcs65\pm0\farcs02$ & $41\fs6305\pm0\fs0004,~43\farcs638\pm0\farcs007$ &  \nodata & \nodata\\
        & 2014/15 & $41.6273\pm0.0005,~43.73\pm0.02$ & $41.6290\pm0.0002,~43.728\pm0.006$ & $41.62760\pm0.00005,~43.732\pm0.002$ & $41.62755\pm0.00004,~43.705\pm0.001$ \\
%
%C (RMS data)  & 2004    & $41.61,~43.4$   & $41.611,~43.35$ & \nodata & \nodata \\
IE  & 2008/09 & $41.880\pm0.003,~44.33\pm0.04$   & $41.885\pm0.005,~44.33\pm0.04  $ &   \nodata & \nodata\\
    & 2014/15 & $41.894\pm0.003,~44.61\pm0.04$    & $41.896\pm0.003,~44.58\pm 0.04$  & $41.897\pm0.002,~44.53\pm0.05$  & $41.894\pm0.002,~44.51\pm0.05$ \\
IW  & 2008/09 & $41.350\pm0.004,~43.09\pm0.04$   & $41.359\pm0.005,~43.01\pm0.03$   &   \nodata & \nodata\\
    & 2014/15 & $41.347\pm0.003,~43.15\pm0.03$    & $41.349\pm0.003,~43.12\pm0.03$   & $41.346\pm0.002,~43.10\pm0.03 $ & $41.350\pm0.002,~43.07\pm0.04$ \\
OE  & 2008/09 & $44.081\pm0.004,~52.2\pm0.2$    & $44.081\pm0.005,~52.0\pm0.1$   &  \nodata & \nodata\\
    & 2014/15 & $44.087\pm0.002 ,~52.39\pm0.06$ & $44.099\pm0.001,~52.23\pm0.05$ & $44.087\pm0.003,~52.36\pm0.08$ & $44.091\pm0.004,~52.39\pm0.07$ \\
OW  & 2008/09 & $39.829\pm0.004,~41.62\pm0.06$    & $39.830\pm0.007,~41.4\pm0.1$  &   \nodata & \nodata\\
    & 2014/15 & $39.802\pm0.002,~41.59\pm0.06$    & $39.764\pm0.001,~41.44\pm0.04$  & $39.750\pm0.001,~41.61\pm0.03$ & $39.753\pm0.001,~41.62\pm0.05$ \\
W$_1$&2008/09 & \nodata                           & $40.18\pm0.02~42.5\pm0.4$      &   \nodata & \nodata\\
     &2014/15 & $40.1207\pm0.007,~42.71\pm0.2$      & $40.145\pm0.002,~42.50\pm0.09$  & $40.139\pm0.001,~42.75\pm0.05$& $40.156\pm0.001,~42.52\pm0.06$ \\
\sidehead{Other sources}
1             & 2014/15 & \nodata         & $42.4297\pm0.002,~30.9\pm0.2$   & $42.427\pm0.002,~30.74\pm0.06$  & $ 42.4238\pm0.0004,~30.98\pm0.05$ \\
2             & 2014/15 &  \nodata        & \nodata                         & $42.850\pm0.001,~36.43\pm0.07$  & $ 42.855\pm0.001,~36.27\pm0.05$   \\
3             & 2014/15 &  \nodata        & $41.710\pm0.002,~48.05\pm0.05$  & $41.678\pm0.002,~47.55\pm0.09$  & $ 41.6832\pm0.0005,~47.95\pm0.02$ \\
\enddata 
\tablecomments{The two numbers given in columns 3 to 6 are the seconds and
  arcsec part of the coordinates of each source. The 5 and 9 GHz positions 
were obtained from the matched resolution images (see Section \ref{sec-pm}).}
\end{deluxetable}

\begin{deluxetable}{lcccc}
\tablewidth{0pc} \tablecolumns{5} \tabletypesize{}
\tablecaption{Flux Densities\label{tab-flux}}
\tablehead{
\colhead{Source} &  \colhead{5 GHz} & \colhead{9 GHz}& \colhead{17 GHz} & \colhead{19 GHz}\\
\colhead{ } &  \colhead{(mJy)} & \colhead{ (mJy) }& \colhead{ (mJy) } & \colhead{(mJy) }
}
\startdata
Central       & $7.6 \pm 0.2$ & $13.4\pm 0.2$ & $20.8 \pm 0.1$ & $23.0 \pm 0.1$ \\
Inner  East   & $6.0 \pm 0.3$ & $6.5 \pm 0.4$ & $4.5 \pm 0.3$ & $4.2 \pm 0.3$ \\
Inner  West   & $6.0 \pm 0.3$ & $6.0 \pm 0.3$ & $4.4 \pm 0.2$ & $3.8 \pm 0.3$ \\ 
Outer  East   & $ 8.6\pm 0.1 $& $8.6\pm 0.1 $ & $ 4.6\pm 0.1 $ & $ 5.1\pm 0.1 $\\
Outer   West  & $ 7.4\pm 0.1 $& $6.7\pm 0.1$  & $ 4.7\pm 0.2 $ & $ 4.1\pm 0.1 $\\
W$_1$\tablenotemark{a} & $ 0.78\pm0.02  $  & $0.61\pm0.02$    & $0.25\pm0.03$ & $0.20\pm0.03$\\
%\sidehead{More Data}  cutinhead{More Data} 
\sidehead{Other sources}
1             & \nodata       & $0.14\pm0.02$ & $0.38\pm0.06$ & $0.23\pm 0.04$\\
2             &  \nodata      & \nodata       & $0.42\pm0.05$ & $0.44\pm0.06$ \\
3             & \nodata       & $0.26\pm0.03$ & $1.2\pm 0.1$  & $0.95\pm0.06$\\
\enddata
%\tablecomments{).}
\tablenotetext{a}{Intensity at peak.}
\end{deluxetable}

The positions listed in Table \ref{tab-pos} for the outer lobes (OE and OW)
correspond to the peak intensity determined using the tool \texttt{maxfit}
in CASA.  To estimate the uncertainty in these positions we explore two
approaches.
In the first approach we note that, although the OE and OW lobe images
have well defined intensity peaks, their morphologies are not
adequately represented by single Gaussians. This is because these
sources are spatially resolved by our observations.
Therefore,  we use two Gaussians to model
each lobe: one of them fits the emission near the intensity peak while the
other reproduces the more extended and resolved emission from the lobe.
For example, the skewed appearance of the OW lobe (suggesting a bow shock)
is better reproduced by an extended Gaussian ($\approx2\farcs5$ FWHM) plus
a compact source ($\approx0\farcs8$ FWHM) shifted  west with respect to
the center of the more extended Gaussian by $\sim1\arcsec$.  We approximate
the uncertainty of the position of the peak as the uncertainty of the
position of the compact Gaussian. This procedure takes into account the
fact that a non-negligible fraction of the peak comes from a diffuse
component, whereas direct application of Equation \eqref{eq-ung} without correction would
underestimate the peak position uncertainty.
The second approach consists of using the clean components and the
synthesized beam to generate a model image for each lobe. To this model
image we add simulated random noise (correlated within the synthesized
beam) at the level indicated in Table \ref{tab-noise}. Then, we use
\texttt{maxfit} to calculate the position of the peak for each of these
simulated images.  The standard deviation of these positions is our
uncertainty estimation of the peak position of each lobe.
% OE 0.006vs0.007,0.08vs0.2; 0.001vs0.002,0.02vs0.06
% OW 0.006vs0.007,0.07vs0.05; 0.001vs0.002,0.02vs0.09
% pdl(0.006/0.007,0.08/0.2,0.001/0.002,0.02/0.06,0.006/0.007,0.07/0.05,0.001/0.002,0.02/0.09)

We find that the uncertainties derived using the first method are larger
than those determined using the second method by about a factor of 2.  We
decide to follow a more conservative approach and use the first method for the position uncertainty of the outer lobes. These uncertainties
are given in Table \ref{tab-pos}.

The flux densities listed in Table \ref{tab-flux} are derived from Gaussian
fittings for sources C, \Ie, and \Iw\ at 5 and 9 GHz. 
At 17 and 19 GHz the inner lobes IE and
IW  
show a compact component, with a 
deconvolved size of
$\sim0\farcs8$, within  significant 
%($\gtrsim30\%$)
extended diffuse emission that cannot be adequately modeled as a
Gaussian. Therefore, we calculate their fluxes integrating over
$5\arcsec\times5\arcsec$ boxes.  We also calculate the flux densities of
the outer lobes at all frequencies by integrating the intensity within
boxes of $9\arcsec\times9\arcsec$. Note that the flux density of the
\Ow\ lobe includes the emission from W$_1$.  Source W$_1$ appears to be
unresolved and blended with the rest of the \Ow\ lobe emission at  9
GHz. Therefore, we report in Table \ref{tab-flux} only its peak intensity.
We also note that source W$_1$ is detected in the 9 GHz 2008/09
images just below the 5-$\sigma$ level, but it was not deemed significant 
and therefore not reported in
\citet{Guzman2010ApJ}.

\begin{figure}
\centering\includegraphics[height=0.85\textheight]{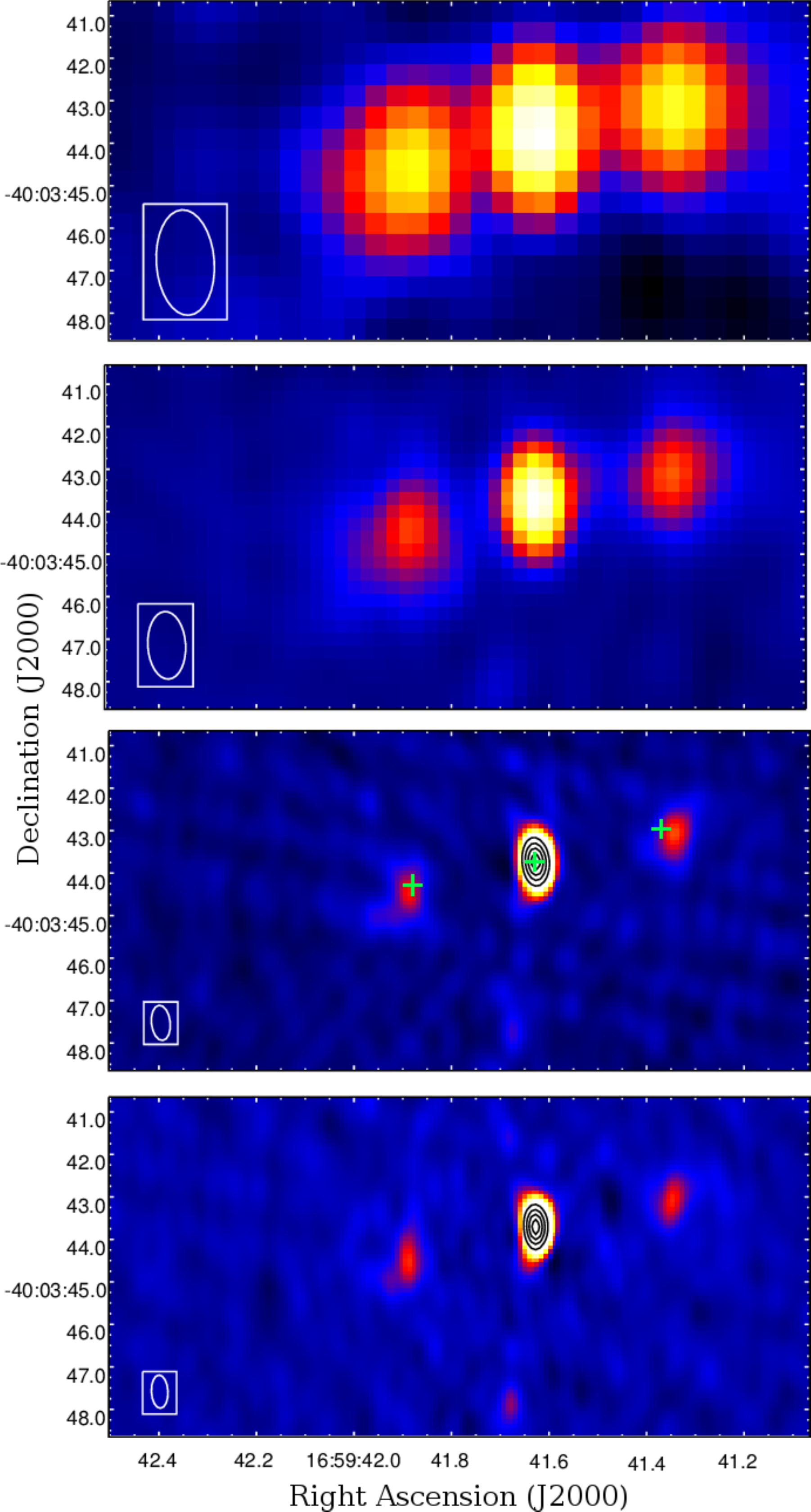}
%\centering\includegraphics[width=\columnwidth]{iiP.pdf}
\caption{\baselineskip=0.64\baselineskip Color images of the radio
  continuum emission detected toward the central source and the two
  inner lobes. From top to bottom: 5, 9, 17, and 19 GHz. Black
  contours in the images at 17 and 19 GHz are drawn at 30, 50, 70, and
  90\% of the peak. Green crosses in the 17 GHz image show the 2011
  positions of sources IE, C, and IW determined by \citet{Purser2016arXiv}.  \label{fig-inn}}
\end{figure}

\begin{figure}
\centering\includegraphics[width=.7\textwidth]{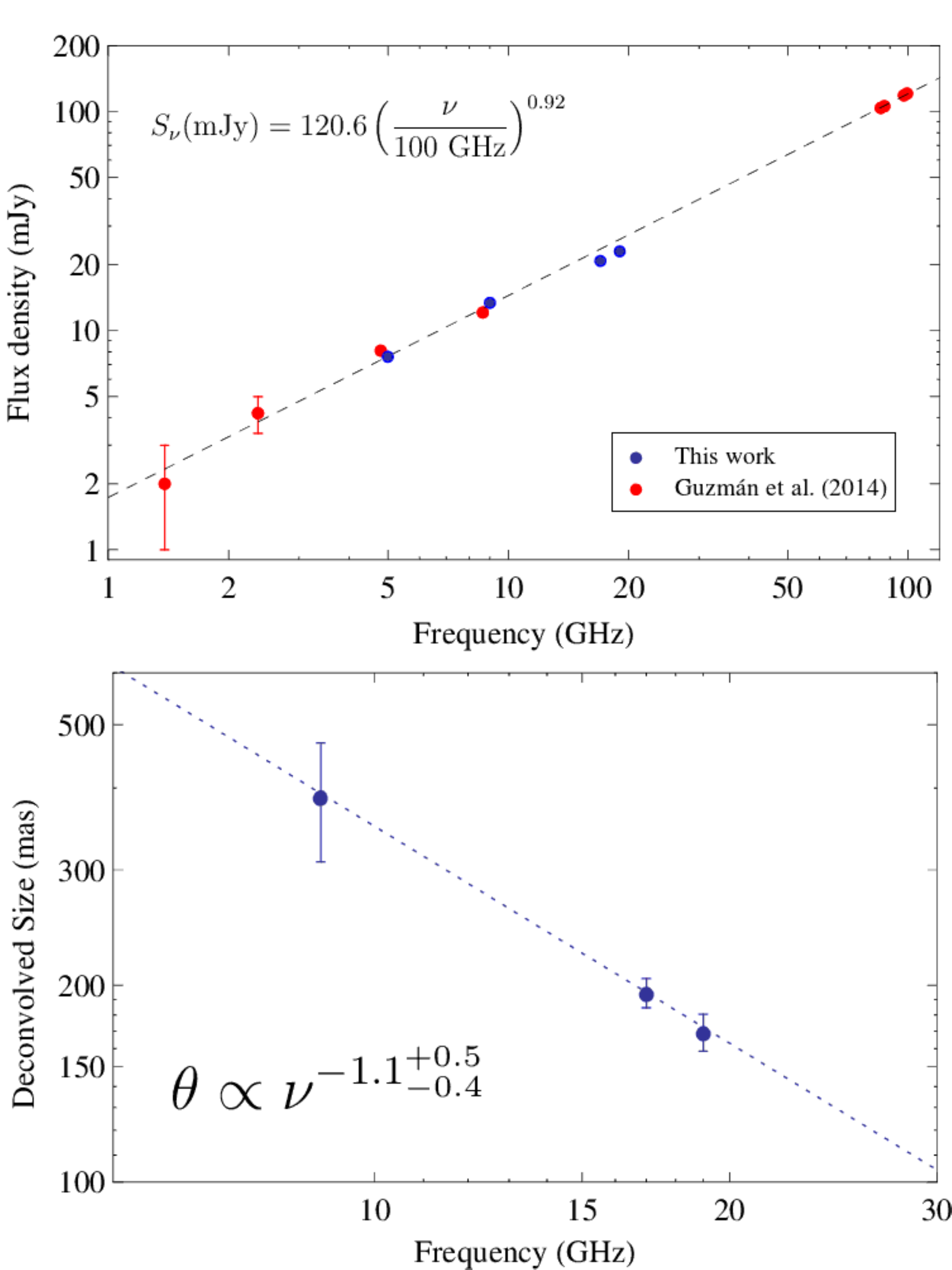}
%\centering\includegraphics[width=\columnwidth]{specs.pdf}
\caption{Top panel: Spectrum of source C, including all previous
  data. The dashed line shows the spectrum derived in
  \citet{Guzman2014ApJ}. Bottom panel: Angular size spectrum of source
  C. The dotted line is the best power law fit to the geometric mean
  of the deconvolved major and minor FWHM axes. \label{fig-specs}}
\end{figure}

Figure \ref{fig-inn} shows the inner region of G345.49+1.47 where sources
C, \Ie, and \Iw\ are clearly distinguishable at the four observed
frequencies.  Source 3 is not detected at 5 GHz.  The deconvolved angular
size of source C, calculated as the geometric mean of the deconvolved major
and minor FWHM axes, are $390\pm80$, $195\pm10$, and $170\pm10$ milliarcsec
(mas) at 9, 17, and 19 GHz, respectively%.
%%
% p.a. de source C are 116, 153 y 143
%%
, while it is unresolved at 5 GHz. While these angular sizes are typically
smaller than half of the synthesized beam, source C is
detected with a signal-to-noise ratio  $\ge20$ at all frequencies,
which helps to reduce the uncertainty of the angular size estimations
\citep[see Equation \ref{eq-ung} and also][sec.\ 5.2.4]{Condon1998AJ}.
These angular sizes are well fitted (residuals $<5$ mas) by a power-law in
frequency with spectral index $-1.1^{+0.5}_{-0.4}$.  Figure \ref{fig-specs}
shows the spectral energy distribution  and angular size versus frequency 
of source C.
%% $13.8\pm0.2(\nu/{\rm 10~GHz})^{0.78\pm0.03}$

%\afterpage{\clearpage}%
{\subsection{Proper Motions}\label{sec-pm}}

Sources C, \Ie, \Iw, \Oe, and \Ow\ are detected in our observations and were detected in the 2008/09 epoch at both 5 and 9 GHz, allowing us to study 
possible changes in their positions.  
Since the resolution of the
2014/15 and 2008/09 images are slightly different, we convolve the images
in order to match their synthesized beams through the procedure described
in detail in Appendix \ref{sec-match}.  The matched synthesized beam of the
5 and 9 GHz images is $2\farcs74\times1\farcs74$ with $\text{P.A.}=2\fdg9$
and $1\farcs68\times1\farcs18$ with $\text{P.A.}=3\fdg3$, respectively.
The pixel size of the 5 and 9 GHz images (the same for both epochs) is
$0\farcs5$ and $0\farcs3$, respectively. These are $\sim1/4$ of the
smallest FWHM of the synthesized beams.  
Table \ref{tab-dip}
lists the displacements measured for each source detected at both epochs
according to the difference between the coordinates given in Table
\ref{tab-pos}. 
\begin{deluxetable}{lccccc}
\tablewidth{0pc} \tablecolumns{6} \tabletypesize{}
\tablecaption{Displacements of the Peak Position\tablenotemark{a}\label{tab-dip}}
\tablehead{\colhead{Source}&\multicolumn{2}{c}{5 GHz}&\colhead{}&\multicolumn{2}{c}{9 GHz}\\
\colhead{}&\colhead{$\Delta$R.A.}&\colhead{$\Delta$Dec.}&\colhead{}&
\colhead{$\Delta$R.A.}&\colhead{$\Delta$Dec.}\\
\colhead{}&\colhead{(\arcsec)}&\colhead{(\arcsec)}&\colhead{}&\colhead{(\arcsec)}&\colhead{(\arcsec)}}
\startdata
C                &  $0.05\pm0.01$ & $-0.08\pm0.03$ && $-0.017\pm0.005$ & $-0.090\pm0.009$ \\
IE               &  $0.07\pm0.05$ & $-0.22\pm0.06$ && $0.14\pm0.07$    & $-0.16\pm0.06$\\
IW               &  $-0.08\pm0.06$& $0.02\pm0.05$  && $-0.10\pm0.07$   & $-0.02\pm0.04$\\
OE               &  $0.02\pm0.05$ & $-0.1\pm0.2$   && $0.22\pm0.06 $    & $-0.1\pm0.1 $\\
OW               &  $-0.36\pm0.05$& $0.11\pm0.08$  && $-0.74\pm0.08$   & $0.1\pm0.1$\\
\enddata
\tablenotetext{a}{The  displacement of source C  has been subtracted from the rest of the sources. }
\end{deluxetable}

\noindent\emph{Central Source.} 
Source C exhibits the smallest displacement, of the order of 100 mas
in the south direction at 9 GHz and in the south-east direction at 5 GHz.  
%$(\Delta{\rm R.A.},\Delta{\rm
%  Dec})=(49\pm13,-80\pm28)$ mas at 5 GHz and $(-17\pm5,-90\pm9)$ mas at 9
%GHz.  
This is somewhat expected due to the nature of source C: being a
HC\ \hii\ region we anticipate possible flux density variations
\citep{DePree2014ApJ} but a relatively stable position.
%% DePree2015tambien 
Note that the displacement of source C is approximately one tenth of
the beamsize and is roughly
aligned with the N-S maximum elongation direction of the synthesized
beams. Therefore, it is possible that source C's apparent proper
motion (PM) signal is an artifact of the observations and not related
with a real displacement.  For example, it could be due to using 
different gain calibrators in the two epochs of observation.
Systematic astrometry shifts between different epochs of the order of
one tenth of the size of the synthesized beam are not rare. For
instance, in a recent PM study presented by \citet{Masque2015ApJ} on HH
80-81 using the VLA, they find shifts of approximately this magnitude
as determined from comparing the positions of extragalactic radio
sources.
Unfortunately, there are no strong extragalactic sources in the field
of G345.49+1.47 which could help us to align both epochs and correct
for a global shift.  If the PM of source C is real, it would imply a
speed of $\sim100$ \kms\ projected in the plane of the sky, which is
unlikely.  However, even in this case, ejecta from the HMYSO would
presumably move at a velocity equal to the source's velocity plus the
ejection velocity. Therefore, to determine the PMs of the lobes respect to
the central source, the PM of source C should be subtracted from the PMs of 
the rest of the sources, being the former due to a systematic 
astrometry error or not.  Hereafter, we subtract source C's
displacement from that of the rest of the sources.

\begin{figure}
\centering\hspace*{-2em}\includegraphics[width=1.1\textwidth]{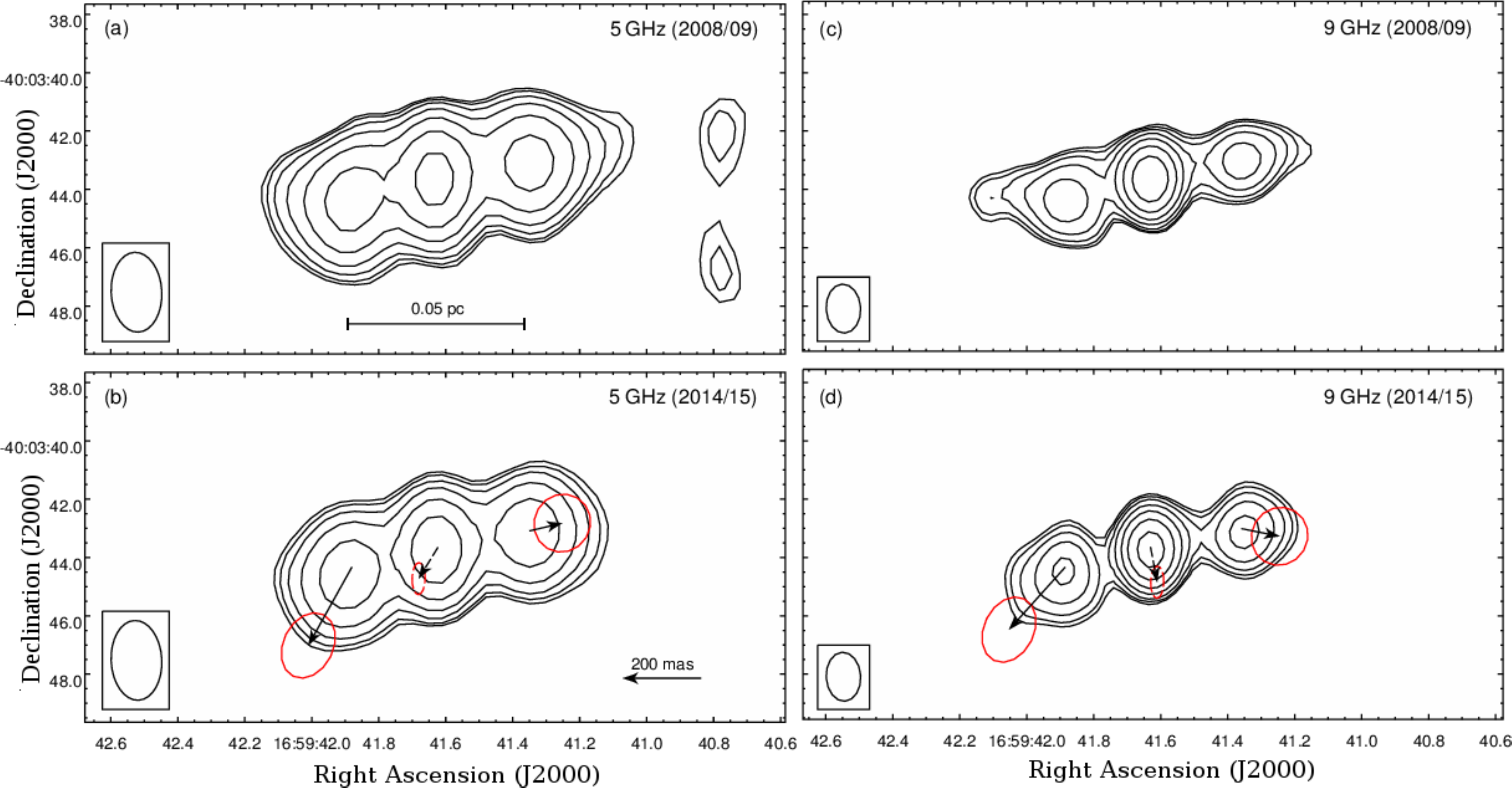}
\caption{\small Source C and inner radio lobes around G345.49+1.47.  Panels
  (a) and (b) show 5 GHz contour maps from the 2008/09 and 2014/15 epochs,
  respectively. Contour levels: $(3+2^i)\times50~\mu$Jy with $i=0\ldots5$.
  Panels (c) and (d) show 9 GHz contour maps from the 2008/09 and 2014/15
  epochs, respectively. Contour levels: $(3+2^i)\times50~\mu$Jy with
  $i=0\ldots6$.  Panels (b) and (d) show the PMs and their
  1-$\sigma$ uncertainty ellipses derived from the 5 and 9 GHz data,
  respectively. PMs of source C --- subtracted from the PMs of sources IE and IW --- are marked using dashed lines. The velocity scale
  of the PM vectors is 100
  \kms\ arcsec$^{-1}$ (assuming a distance of 1.7 kpc and a 6 yr timescale).\label{fig-pminn}}
\end{figure}
\noindent\emph{Inner Lobes.} Figure \ref{fig-pminn} compares the 5 and 9 GHz images taken in both epochs
toward the inner sources (C, IE, and IW). 
The arrows in 
the lower panels represent the PM signals of the inner lobes and of source
C, given in Table \ref{tab-dip}.  The ellipses show the 1-$\sigma$
uncertainty regions. We find that the IE lobe has a PM equivalent to a
velocity in the plane of the sky of $\sim300$ \kms, with a more clear
detection than that of the IW lobe. The PM signal of the IW lobe is weaker,
being close to 2-$\sigma$ at 9 GHz and only slightly above $1$-$\sigma$ at 5
GHz.   Figure \ref{fig-inn}c  also shows 
the peak position of the inner lobes determined from observations at 17 GHz carried out during 2011 \citep{Purser2016arXiv}. 
The clear displacement observed between our data and that taken independently provides further support for the PM of the inner lobes.
%% The 1-$\sigma$ uncertainty ellipses are defined as enclosing $68.27\%$ (a
%% $\pm\sigma$ in a 1D-Gaussian) of the probability. In practice, if
%% $\sigma_M$ and $\sigma_m$ are the two standard deviations  along the principal axes 
%% of a 2D-Gaussian 
%% representing the position uncertainty, then the 1-$\sigma$ ellipse
%% is defined with the same orientation and semi-axes equal to
%% $\sqrt{-2\ln\erfc\left(2^{-1/2}\right)}\sigma_M\approx1.52\sigma_M$ and
%% $1.52\sigma_m$, where $\erfc$ is the complementary error
%% function. That is, the size of the 1-$\sigma$ ellipses drawn in Figure \ref{fig-pminn}
%% are \emph{larger}   (by a 52\%) than the uncertainty ellipses that have
%% semi-axes equal to $\sigma_M$ and $\sigma_m$.

\begin{figure}
\centering\includegraphics[width=\textwidth]{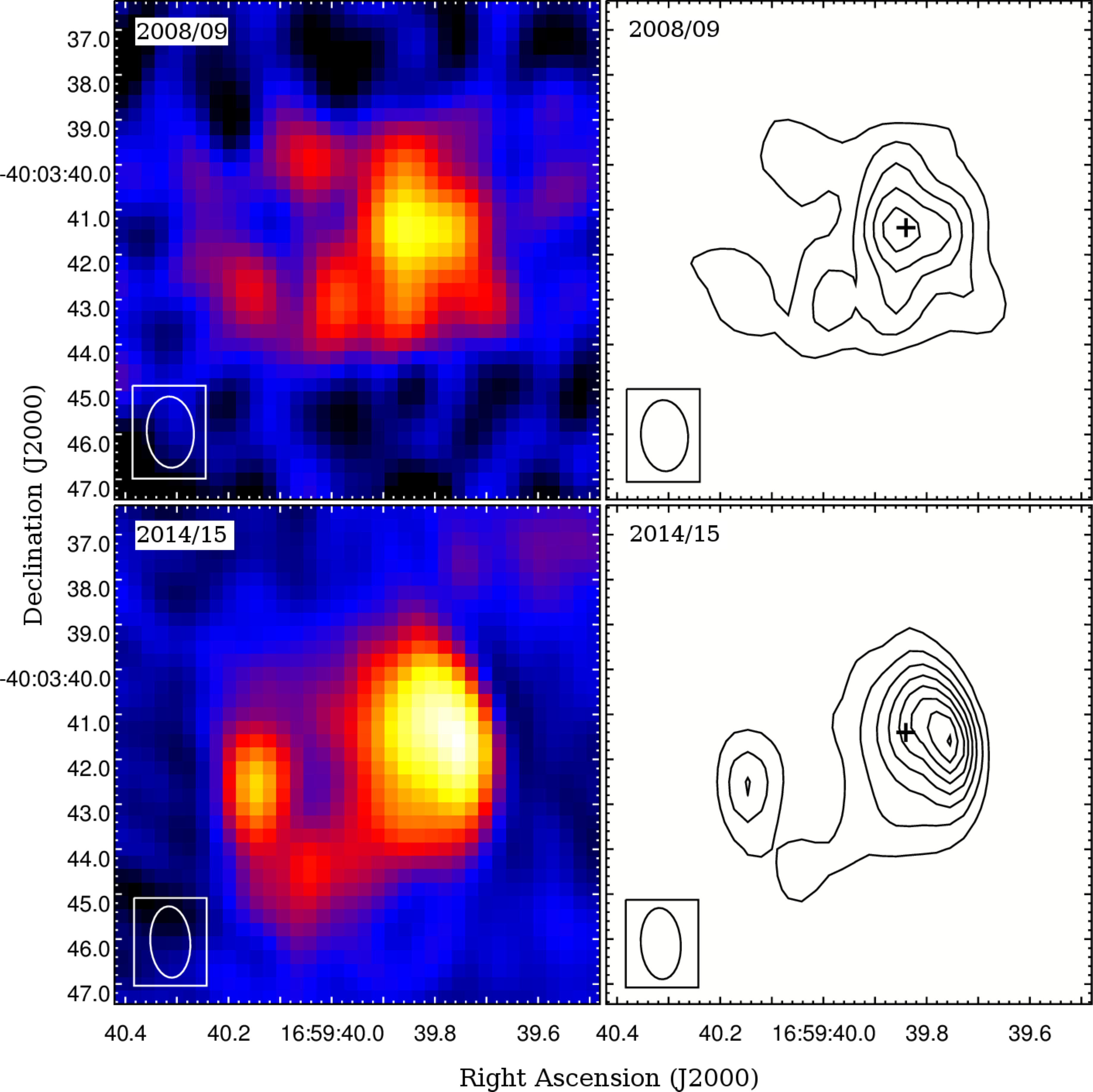}
\caption{\small Color images (left) and contour maps (right) of the 9 GHz
  emission from the \Ow\ lobe including knot W$_1$. Top: 2008/09 epoch. Bottom: 2014/15
  epoch.  The cross in the right panels indicate the position of the peak
  emission observed during 2008/09 epoch. Contour levels are: from 1 to 5
  $\times5\sigma$ with $\sigma=75~\mu$Jy in the top right panel and from 1
  to 8 $\times 12\sigma$ with $\sigma=17~\mu$Jy in the bottom right
  panel.\label{fig-pmow}}
\end{figure}
\begin{figure}
\centering\includegraphics[width=\textwidth]{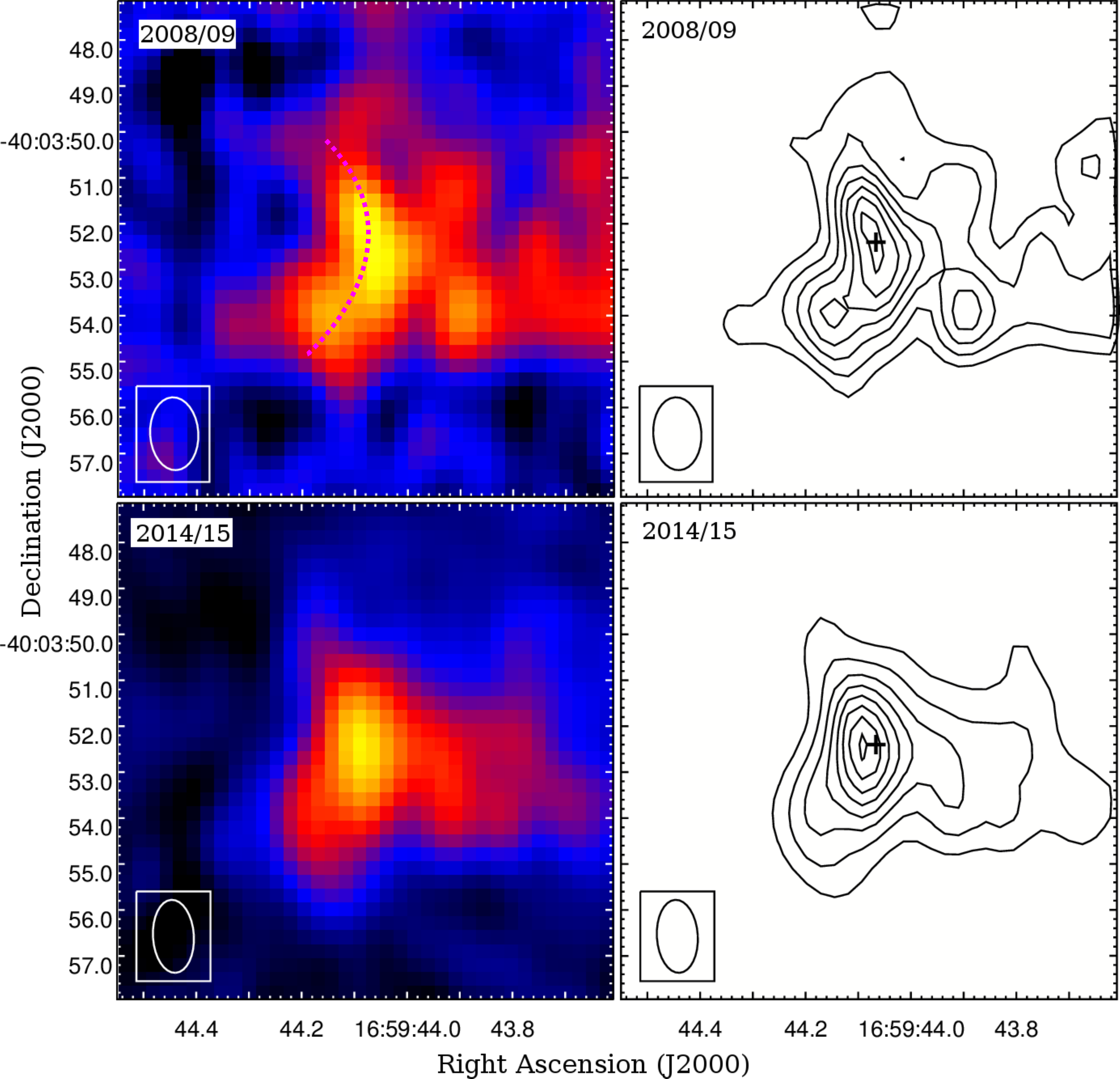}
\caption{ \small Color images (left) and contour maps (right) of the 9
  GHz emission from the \Oe\ lobe. Top: 2008/09 epoch. Bottom: 2014/15
  epoch.  The cross in the right panels show the position of the peak
  emission observed during 2008/09 epoch.  The contour levels are:
  from 1 to $7\times 2\sigma$ with $\sigma=75~\mu$Jy in the top right
  panel and from 1 to $8\times 10\sigma$ with $\sigma=17~\mu$Jy in the
  bottom right panel. The dashed magenta curve in the top-left panel
  indicates the arc-like shape of the emission discussed in Section
  \ref{sec-pm}.\label{fig-pmoe}}
\end{figure}
\noindent\emph{Outer Lobes.}  The morphology of the external lobes OE
and OW is considerably more complex than that of the inner lobes.
Images and contour maps of the 9 GHz emission from the OW and OE lobes
in the two epochs are shown, respectively, in Figures \ref{fig-pmow}
and \ref{fig-pmoe}.  We emphasize three characteristics of the OW
lobe's emission: (i) the shape of the main
part of the lobe resembles a detached or bow-shaped shock, (ii) the
peak of the emission has displaced away from the central source by
$\sim1\arcsec$, and (iii) the presence of the secondary knot W$_1$ (see Figure \ref{fig-pmow}).
As shown in Figure \ref{fig-pmoe}, the OE lobe's shape is better
defined in the 2014/15 images respect to the 2008/09 images.  While
the displacement of the peak position of the OW lobe is readily
patent, that of the OE lobe is less clear.  We note that the \Oe\ lobe
is likely interacting with a dense gas structure which appears as a
dark patch in the IR three-color image of Figure
\ref{fig-pres}. Furthermore, the shape of the OE emission at 9 GHz
seems to form an arc curving away from source C.  The dashed magenta
line in the top left panel of Figure \ref{fig-pmoe} shows
schematically the location of this arc.  A possible interpretation is
that it corresponds to a detached shock produced by the interaction of
jet material with a stationary and dense cloudlet, as in the model of
\citet{Schwartz1978ApJ}.

The rich structure of the OE and OW lobes introduces an
additional difficulty when measuring displacements between two epochs.
Therefore, we use two methods to measure their PMs: (i) by
calculating the displacement of the peak of the emission, whose
results are given in Table \ref{tab-dip}, and (ii) by maximizing the
cross-correlation respect to displacements between the two images.
The cross-correlation method has been applied frequently in PM studies
of extended emission associated with low-mass protostellar jets
\citep[e.g.,][]{Reipurth1996AA,Raga2012ApJ}.
%it is more reliable
%compared to directly estimating the PM by tracking selected
%features in the images \citep{Raga2013AJ}.
The cross-correlation between two images $I_1$ and $I_2$ in a region
$\mathcal{R}$ is defined as
\begin{equation}\label{eq-cor}
\bm{\rho}_{12}(dx,dy)=\frac{\sum_{(x,y)\in\mathcal{R}}I_1(x,y)I_2(x-dx,y-dy)}{\left(\sum_{\mathcal{R}}I_1^2(x,y)\times\sum_{\mathcal{R}}I_2^2(x-dx,y-dy)\right)^{1/2}}~~,
\end{equation}
where the sums run over all pixels with coordinates $(x,y)$ in
$\mathcal{R}$ and $(dx,dy)$ is the displacement of image $I_2$ with respect to
$I_1$. Clearly we always have $|\bm{\rho}_{12}|\le1$. If there is no
noise and $\bm{\rho}_{12}(\tilde{dx},\tilde{dy})=1$, then the morphology of
the emission inside $\mathcal{R}$ is preserved exactly between $I_1$ and  $I_2$, 
and $I_1(x,y)=\text{constant}\times I_2(x-\tilde{dx},y-\tilde{dy})$. Therefore,
in addition to the best-fit displacement, the
maximum value attained by $\bm{\rho}_{12}$ is a measure of the 
degree of similarity
between  the two epochs. We implement the cross-correlation
and its maximization using PDL\footnote{http://pdl.perl.org/} and the
package Minuit.  To evaluate the cross-correlation at sub-pixel size
accuracy we linearly interpolate the images using the task
\texttt{interpND} within PDL.

\begin{deluxetable}{lcccccc}
\tablewidth{0pc} \tablecolumns{7} \tabletypesize{\small}
\tablecaption{Displacements of Extended Lobes from Cross-Correlation Maximization\label{tab-dic}}
\tablehead{\colhead{Source}&\multicolumn{3}{c}{5 GHz}&\multicolumn{3}{c}{9 GHz}\\
\colhead{}&\colhead{$\Delta$R.A.}&\colhead{$\Delta$Dec.}&\colhead{$\bm{\rho}_\text{max}$}&
\colhead{$\Delta$R.A.}&\colhead{$\Delta$Dec.}&\colhead{$\bm{\rho}_\text{max}$}\\
\colhead{}&\colhead{(\arcsec)}&\colhead{(\arcsec)}&\colhead{}&\colhead{(\arcsec)}
&\colhead{(\arcsec)}&\colhead{}}
\startdata
OE & $ 0.08\pm0.07$ & $-0.06\pm0.06$    & 0.94 & $ 0.22 \pm0.08 $ & $ 0.19 \pm0.08 $ & 0.93\\
OW & $-0.40\pm0.05$ & $0.10\pm0.08$     & 0.97 & $-0.38 \pm0.07 $ & $-0.06  \pm0.08 $ & 0.86 \\
\enddata
\end{deluxetable}
Table \ref{tab-dic} gives the displacements calculated by maximization of
the cross-correlation toward the OE and OW lobes.  Columns 2 and 3 give the R.A.\ and declination
displacements for the 5 GHz images and column 4 lists the value of
$\bm{\rho}_{\rm max}$. Columns 5--7 give these same quantities but for the
9 GHz images.  To determine the uncertainty of each PM signal we
construct synthetic images of the OE and OW lobes convolving the clean
components of these regions (obtained from the CASA task \texttt{clean})
and adding simulated Gaussian noise at the level given by Table
\ref{tab-noise} (correlated spatially according to the synthesized
beams). We introduce artificial PM signals to each simulation by
shifting one of the synthetic images by a random vector.  We recover the
PM signal through cross-correlation maximization and calculate
its difference with the one introduced artificially. This difference
$(\Delta \alpha,\Delta\delta)$ gives us an estimation of the uncertainty of
the cross-correlation method. Repeating this process $N$ times generates a
$N\times2$ (rows $\times$ columns) matrix of differences, which we denote
$\bm{D}$. The correlation matrix $\bm{D}^{\rm T}\bm{D}$ define the
uncertainty ellipses of the cross-correlation method. For simplicity, in
Table \ref{tab-dic} we only give the projections of the 1-$\sigma$
uncertainties onto the R.A.\ and declination axes. We perform this process
on each of the OE and OW lobes at 5 and at 9 GHz using $N=300$. Larger
values of $N$ do not change the results appreciably.

\begin{figure}
\centering\includegraphics[width=0.92\textwidth]{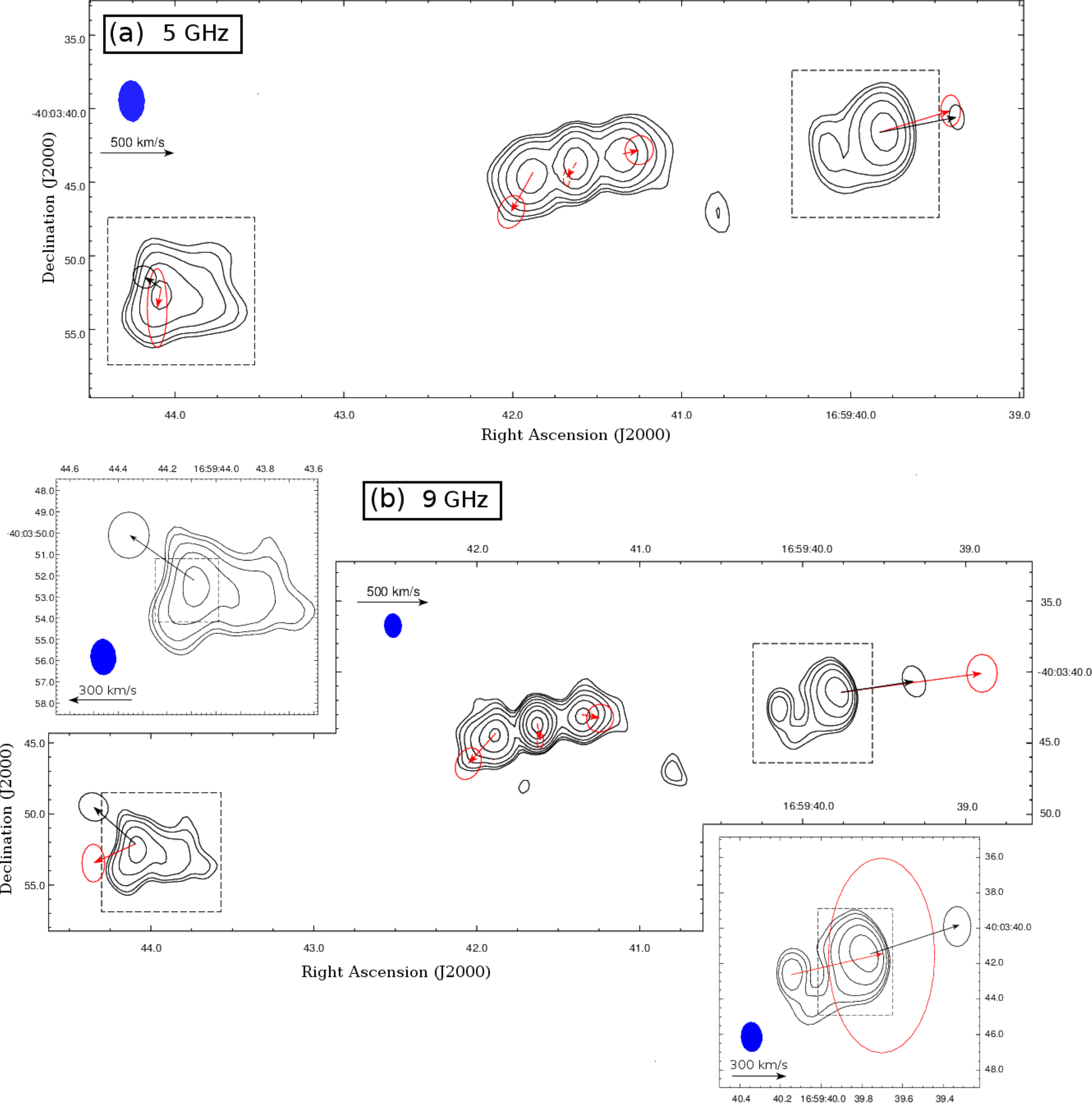}
%\centering\includegraphics[width=\textwidth]{pms.pdf}
\caption{\small  \baselineskip=0.64\baselineskip Contour maps and PMs of
  G345.49+1.47 at 5 and 9 GHz.  Blue filled ellipses display the beam. Red
  vectors and ellipses indicate respectively the PMs and the
  uncertainty regions derived from displacements of the peak.  The dashed
  line boxes and black vectors indicate the regions used for calculating
  the cross-correlation and the derived PMs, respectively.  The
  velocity scale of the vectors is 1\arcsec\ per 100 \kms.  For sources IE,
  IW, and C the PMs are the same as in Figure
  \ref{fig-pminn}. PM of source C has been subtracted from
  that of the rest of the sources, at each frequency.  Panel (a): 5 GHz
  data.  Contour levels: $2+2^i\times50 \mu$Jy, $i=1\ldots6$.  Panels (b):
  9 GHz data. Contour levels: $2+2^i\times50 \mu$Jy, $i=0\ldots8$. Insets
  at the top left and bottom right corners of Panel (b) show zooms of the
  OE and OW lobes, respectively, and show the results of cross-correlation
  PMs over more restricted regions around the peak of each
  lobe. The bottom right inset shows the PM of
  W$_1$.\label{fig-pms}}
\end{figure}

In Figure \ref{fig-pms} we summarize the PMs measured independently in
the 9 and 5 GHz images. Specifically, plotted are the velocities
projected in the plane of the sky assuming a time scale of 6 yr and a
distance of 1.7 kpc. Red vectors indicate velocities derived from the
displacement of the peak (Table \ref{tab-dip}) and black vectors those
derived from maximizing cross-correlations (Table \ref{tab-dic}). The
dashed line boxes indicate the regions used to calculate the
cross-correlations.  The bottom right inset of panel (b) in Figure
\ref{fig-pms} shows the PM of the main OW lobe (calculated using
cross-correlation maximization) and of the W$_1$ knot (calculated as
the difference between the positions given in Table \ref{tab-pos}).
All PMs toward the OW lobe at 9 and 5 GHz are consistent, except that
the PM of the peak of the OW lobe at 9 GHz is a factor of $\sim2$
larger than the rest of the measurents toward OW.

\begin{figure}
\centering\includegraphics[width=\textwidth]{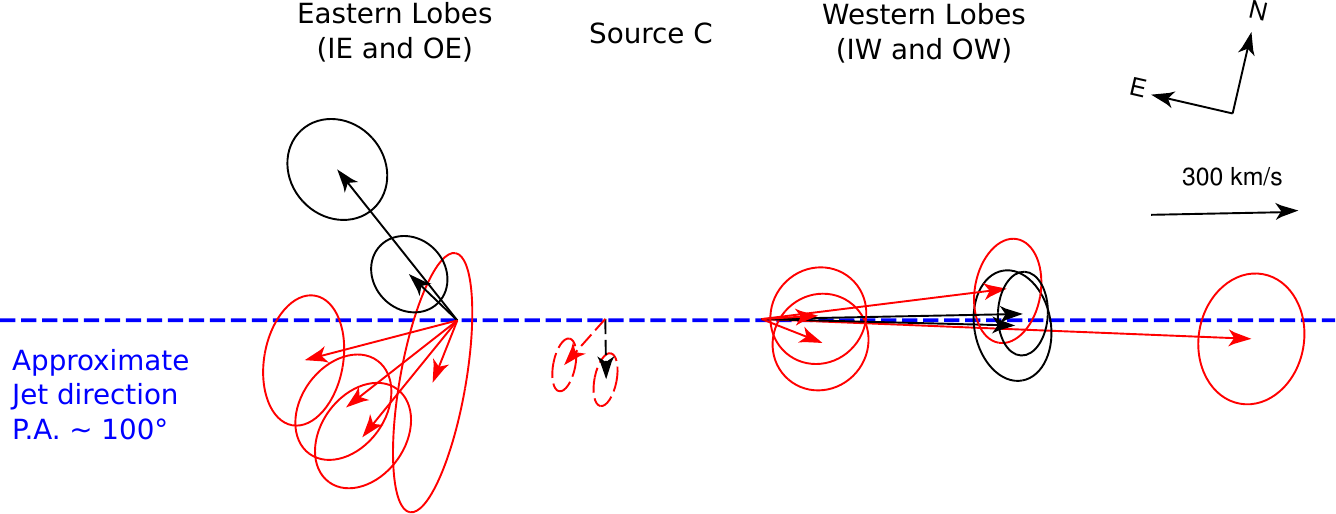}
\caption{Diagram showing the direction and magnitude of the PMs. The colors
  of the vectors and ellipses follow the same convention as in Figure
  \ref{fig-pms}. The orientation of the diagram and the scale of the
  vectors and ellipses are given in the top right corner. The vectors are
  assorted in three groups: Eastern lobes, Source C, and Western lobes. In
  each of these groups the PM vectors have the same
  origin. \label{fig-fle}}
\end{figure}

We further
note that the directions of the PMs of the OW and IW lobes 
lie along the approximate jet orientation pointing to the west. 
Figure \ref{fig-fle}
shows this more clearly, by displaying all PM vectors of the western lobes
(OW and IW) with the same origin. 
%As noted before, the PM of the IW lobe is
%significatively smaller than that of OW.  On the eastern side, the PM of the
%OE lobe seems to be more affected by its association with more extended
%emission than the OW lobe. 
Figure \ref{fig-fle} also shows the PMs of the eastern lobes (IE and
OE). The direction of the eastern lobes PMs have considerable more
dispersion, although they also apparently scatter around the direction
of the jet. One characteristic of the OE PMs which repeats at 5 and 9 GHz
is the widely different direction of the PM as determined from the
peak compared with that determined from cross correlations. The latter
 seems to have a strong component toward the north.

Another characteristic of the outer lobes, evident in Figures
\ref{fig-pmow} and \ref{fig-pmoe}, is that their morphology has
changed somewhat between the 2008/09 and 2014/15 epochs. These changes
can produce a PM signal not related with true displacements and they
may account for the differences between the PMs determined using the
two methods (peak position comparison and cross-correlations
maximization).  As mentioned before, the value of $\bm{\rho}_{\rm
  max}$ is a quantitative measure of the image similarity between the
epochs.  From Table \ref{tab-dic}, we note that in all cases
$\bm{\rho}_{\rm max}$ is close to 1 (columns 4 and 7), with the values
at 5 GHz being larger than those measured at 9 GHz.  The latter is
likely due to the angular resolution at 5 GHz being lower that at 9
GHz, which helps to homogenize the morphology of the sources (at very
low angular resolution, they would become two unresolved point
sources).  To determine how much does $\bm{\rho}_{\rm max}$ decreases
due to random noise, we calculate it between identical images of the
OE and OW lobes but with the addition of simulated noise. We obtain
values ranging between $0.98$--$0.99$ and $0.94$--$0.98$ for the
simulated 5 and 9 GHz data, respectively, showing that the measured
values of $\bm{\rho}_{\rm max}$ are not fully explained by noise.

\begin{figure}
\centering\includegraphics[width=\textwidth]{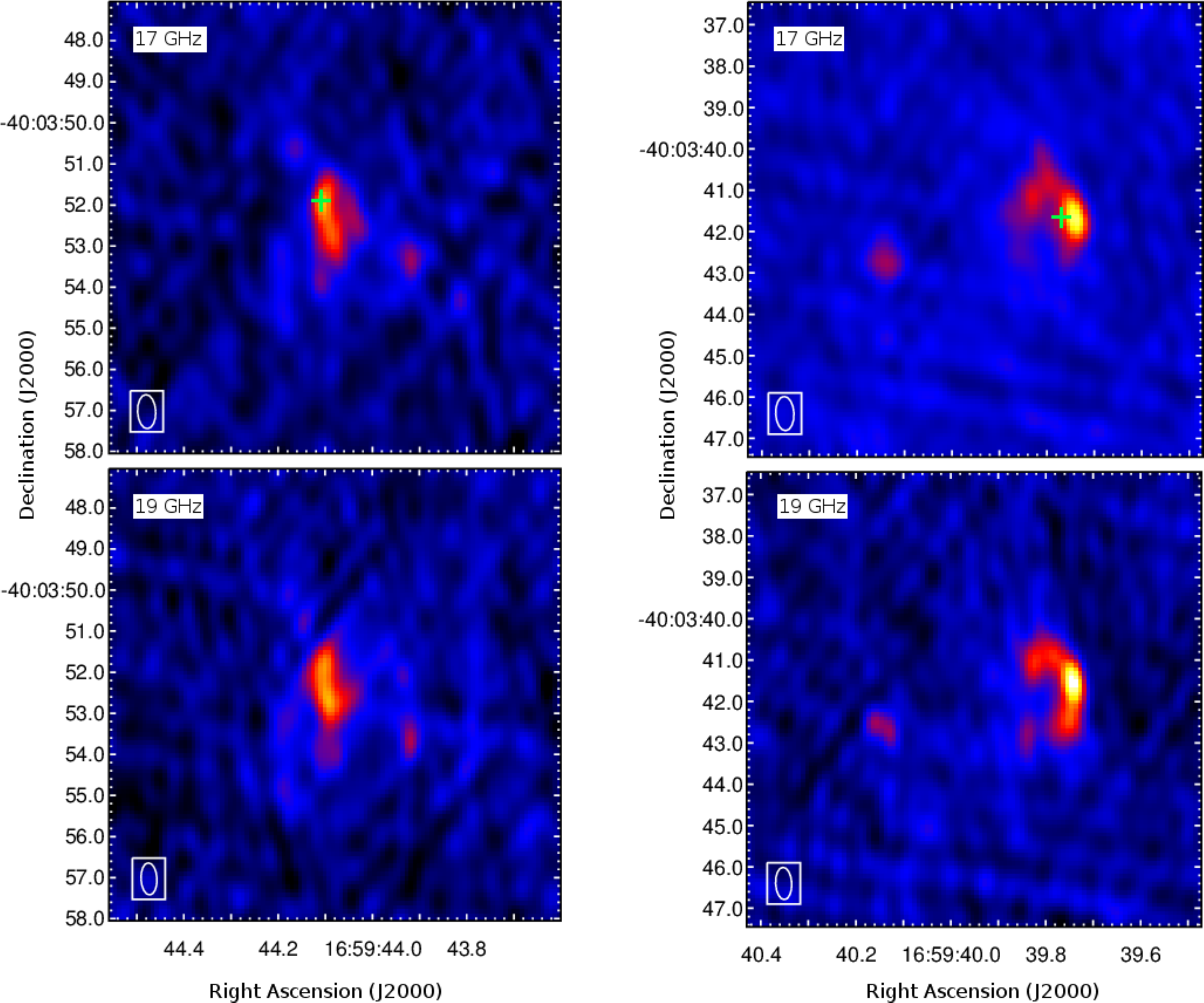}
\caption{Radio continuum images of the outer lobes. Top panels: 17
  GHz. Green crosses mark the 2011 peak position of OE and OW
  determined by \citet{Purser2016arXiv}. Bottom panels: 19 GHz.  Left
  panels: \Oe\ lobes, right panels: \Ow\ lobes. \label{fig-o1719}}
\end{figure}

We conclude that, while the high values of $\bm{\rho}_{\rm max}$
indicate that the bulk of the emission from the outer lobes has kept
its morphology, there are still statistically significant changes in
shape between both epochs.  In consequence, the differences in
morphology between the two epochs for the outer lobes make necessary
to support their interpretation as PMs with additional evidence, for
example, the bipolar anisotropy in the direction of the PMs  shown
in Figure \ref{fig-fle}.

Finally, Figure \ref{fig-o1719} shows images of the outer lobes at 17 and
19 GHz (2014/15 epoch only).  In part due to the likely filtering out of
extended emission by the interferometer, the morphologies of lobes \Ow\ and
\Oe\ are much more similar at these frequencies than they are at 9 or 5
GHz: both outer lobes display an arcuate morphology curved in all cases
toward source C.  Also plotted in Figure \ref{fig-o1719} are the peak
position of the outer lobes determined from observations at the same
frequency carried out in 2011 \citep{Purser2016arXiv}, confirming the PM of the OW lobe. The observed displacement
of the OE lobe, on the other hand, seems similar to the PM determined from
the cross-correlation method at 9 GHz, that is, rather perpendicular to the
jet direction.

%{\clearpage}%%
{\section{DISCUSSION}\label{sec-dis}}

We adopt some of the nomenclature used by earlier studies 
and  denote as G345.49+1.47 the HMYSO which
dominates the luminosity of the massive clump identified with IRAS
16562$-$3959.  Based on its luminosity, the HMSYO has $\approx15\Msun$ while
 the mass of the IRAS 16562$-$3959 massive molecular clump is
$\approx1000\Msun$.  In this work and in \citet{Guzman2010ApJ}, we identify
five radio sources associated with G345.49+1.47: four lobes and source C. 
The position of the 
latter is coincident,
within 0\farcs2, with the position of the near-IR source
identified in band K$_S$ in Section \ref{sec-res}. It is also consistent, within 0\farcs5, with the position of  the HMYSO given by the RMS Survey
\citep[][]{Lumsden2013ApJ}. Source
C was also identified at 3 mm by \citet[their source 10]{Guzman2014ApJ} .

In previous works \citep{Guzman2010ApJ,Guzman2011ApJ,Guzman2014ApJ} the
radio emission between 1 and 100 GHz arising from source C has been
referred to as a ``jet,'' nomenclature typically used to denote highly
collimated streams of partially ionized, high-velocity ($>100$ \kms) gas,
where the ionization is shock induced \citep[see
  however,][]{Tanaka2016ApJ}.  The HRL observations of
\citet{Guzman2014ApJ} indicate, however, that the radio emission from
source C corresponds to a HC \hii\ region, possibly expanding, but at a
much lower velocity.  The ionized gas is likely being excited by UV
photons arising from G345.49+1.47. Thus, the ``jet'' nomenclature for
source C is inappropriate.
 
\subsection{A Fast Protostellar Jet from a HC \hii\ Region}
%R -> 5.2 Rsol, M -> 15 Msol, Vesc=1050 kms

In star formation studies, mechanisms involving magnetic fields threading
an accretion disk seem to better explain the  acceleration
and collimation of jets.  Magneto-centrifugal acceleration sufficiently close to the
star, for example, can produce velocities comparable to  the escape speed. In addition, the magnetic field can collimate the ejected
material close to the accreting star and explain the narrow cross section
of jets.
Other explanations of jet formation face several theoretical and
observational problems  \citep{Cabrit2007LNP}:
\begin{itemize}
\item{Purely hydrodynamic mechanisms (e.g., Parker type winds,
  hydrodynamical nozzles) are able to accelerate the material only up to a
  few times the sound speed. Therefore, to explain velocities in excess of
  100 \kms, significant amounts of $\sim10^6$ K gas is needed, which is not
  observed. On the other hand, acceleration by stellar radiation has proven
  insufficient to explain the momentum deposited in the associated
  molecular outflows.}
\item{In the low-mass case, collimation of the jet occurs relatively
  near the star, where the jet pressure is still too large to be
  confined by the ambient material. Recent observational evidence
  suggests that the jet collimation in high-mass stars also occurs
  near the HMYSO
  \citep{Greenhill2013ApJ,Carrasco-Gonzalez2015Sci,Caratti-o-Garatti2016AA}. It may be possible
that the larger
  amount of material in high-mass star cores could help to confine the jet, but better angular resolution observations are still necessary to solve this issue.
%% In the case of
%%   G345.49+1.47, even accounting for the dense envelope material and adding
%%   the ram pressure of the collapsing clump \citep{Guzman2011ApJ}, at 300 AU
%%   from the star the ambient pressure still fails --- by an order of
%%   magnitude --- to collimate a jet  with the momentum flux of the
%%   molecular outflow \citep{Guzman2012thesis}. 
%% Note that the estimations of
%%   the mass outflow and momentum rates of the G345.49+1.47 ``jet'' made in
%%   \citet{Guzman2010ApJ} assumed  that the ionized gas from
%%   source C was moving at $\sim300$ \kms\ and therefore they need to be revised.
}
\end{itemize} 

The relevance of finding collimated jets toward HMYSOs is that they
signpost disk accretion. Jets moving at velocities comparable to the escape
velocity further indicate that the ejection mechanism is linked to
accretion onto the central HMYSO, in a similar fashion as in low-mass stars
\citep{Shu1987ARA&A,Li2014PrPl}.  PMs of radio lobes tracing jets have been
measured in high-mass \citep{Marti1998ApJ,Curiel2006ApJ,Rodriguez2008AJ,Carrasco-Gonzalez2010AJ} and
low-mass YSOs \citep{Curiel1993ApJ,Rodriguez2000AJ}. In G345.49+1.47, the evidence
for the radio lobes  being excited by an
underlying fast jet can be summarized as follows:
\begin{enumerate}
\item Most of the lobes display PMs. The magnitude of the PMs ranges between fast and
  highly significant signals for the OW lobe ($\gtrsim500$ \kms) to lower tangential velocities
   ($100\text{--}300$ \kms) for the IE lobe and  OE lobe at 9 GHz. 
\item The lobes are  aligned roughly in the East-West direction
  ($\text{P.A.}\approx100\arcdeg$). As shown in Figure \ref{fig-pms} and
  \ref{fig-fle}, there is an evident anisotropy in the distribution of
  directions of the PMs: eastern lobes move roughly toward eastern directions
  (although with a large scatter) and western lobes move westerly. Both
  groups of lobes at each side of G345.49+1.47 recede from it, as expected
  in the case of a protostellar jet. This anisotropy is confirmed even by
  PMs detected with a low signal-to-noise ratio, such as those of the IW lobe. 
\item The \Ow\ lobe displays an unmistakably bow shock shape with an
  orientation consistent with moving away from G345.49+1.47. The shape of
  the \Oe\ lobe  at 9 and 5 GHz is more complex, but its morphology
  at 17 and 19 GHz suggests that it might also be a bow shock receding from
  G345.49+1.47. 
  \item A comparison of our 17 GHz images with independent data taken
    during April 2011 by \citet{Purser2016arXiv} confirms the PM
    measurements reported in this work for the IE, IW, and OW lobes. In
    addition, evidence of fast shocks in the form of extended X-ray
    emission was found recently toward the OE lobe (V.\ A.\ Montes 2016,
    personal communication). 
\end{enumerate}

Based on the displacements in a timescale of 6 yr observed at 9 GHz,
we determine that the PMs of the \Oe, \Ie, \Iw, and \Ow\ lobes are
$48\pm13$, $35\pm10$, $17\pm11$, and $64\pm12$ mas yr$^{-1}$,
respectively.  For the OW lobe, we use the PM calculated by the
cross-correlation method because it is in better agreement with the
one calculated at 5 GHz.  Assuming a distance of 1.7 kpc, these PMs
correspond to velocities of $390\pm100$, $280\pm80$, $140\pm90$, and $520\pm100$ \kms\ in
the plane of the sky, respectively.  The dynamical time of the lobes,
calculated as the angular distance to source C divided by the PM, are
$600\pm200$, $90\pm30$, $200\pm100$, and $330\pm60$ yr for the \Oe,
\Ie, \Iw, and \Ow\ lobes, respectively. In principle, the underlying jet
has been active (although likely with bursts of activity) during an interval
spanning at least the range of  dynamical times.

The relevance of G345.49+1.47 is that it shows that a collimated
and fast jet can be generated from a HMYSO which is already
producing ionizing radiation.  In other cases of jets
associated with HMYSOs with luminosities $>30,000$ \Lsun\ like
G343.1262$-$00.0620
\citep{Rodriguez2008AJ}, IRAS 13481$-$6124
\citep{Kraus2010Nat,Caratti-o-Garatti2015AA}, or G35.2$-$0.7N
\citep{Gibb2003MNRAS}, it is not clear that the HMYSO is already
producing ionizing photons or that the central radio source
corresponds to a photoionized region.
Theoretical work describing disk accretion and jet ejection under the
influence of ionizing radiation has been rather
scarce. \citet{Tan2003arXiv} and \citet{Tanaka2016ApJ} investigated
the ionization structure of a parameterized model of a collapsing
massive core (including a disk and a disk wind) under UV photon
injection.  However, there are differences between our physical
interpretation and theirs. In \citet{Tanaka2016ApJ}, the ionized
magnetically accelerated disk wind is the HC \hii\ region, predicting
HRLs which are wider than the ones observed toward G345.49+1.47.  In
\citet{Guzman2014ApJ}, on the other hand, we suggest that the
expansion of the HC \hii\ region is hydrodynamical (instead of
magneto-centrifugal). We modeled the observed HRL widths with a
combination of pressure and opacity broadening, and no contribution
from bulk motions. The radio continuum and HRL emission of the jet
itself is, in our interpretation, not detected.

\subsection{Nature of the Radio Emission}

\subsubsection{Source C and the Lobes}

From the observations presented in this work we provide further support to
the simple model of source C, as consisting of a biconical HC \hii\ region
\citep{Guzman2014ApJ}. The flux density spectrum of source C (Figure
\ref{fig-specs}, top panel) follows well the power law fitted in
\citet{Guzman2014ApJ}, except perhaps by a slight decrement of the observed
17 and 19 GHz fluxes with respect to the fit.  This decrement is unlikely due to
a hypothetical optically thin turn-off since, if this were the case, the
HRL fluxes of source C would be inconsistent with the continuum flux
measured at 3 mm \citep{Guzman2014ApJ}.  It is also apparent from Figure
\ref{fig-specs} that the spectral index of the fluxes at 17 and 19 GHz is
similar to the overall power-law fit.  Hence, we attribute the offset either to a calibration error that produces an underestimation of $\sim12\%$ in the
flux of source C, or to intrinsic flux variations  at 17 and 19 GHz \citep{Galvan-Madrid2011MNRAS}.

In the lower panel of Figure \ref{fig-specs} we plot the observed
deconvolved size of source C as a function of frequency, showing a
clear decrease with frequency. A power-law fit to the data gives an
spectral index of $-1.1^{+0.5}_{-0.4}$.  The apparent size of a
biconical HC \hii\ region at a given frequency depends on the distance
from the young star at which the free-free emission becomes optically
thin.  Using the notation of \citet{Reynolds1986ApJ}, this distance
depends on frequency as $\nu^{2.1/q_\tau}$. Considering that the
aperture of the wind cavity is proportional to $r^\epsilon$, where $r$
is the distance to the central young star ($\epsilon=1$ being a
conical wind), $q_\tau=-3\epsilon$ in the case of isothermal and
non-accelerating winds.  Therefore, the angular size of the source
follows the relation $\theta\propto\nu^{-0.7/\epsilon}$. We derive
$\epsilon=0.6^{+0.6}_{-0.1}$ which is compatible with the range of
parameters given by \citet[Eq.\ (16)]{Guzman2014ApJ}. It is also
compatible, within the uncertainty, with a conical wind.

The spectral energy distribution of the lobes  seems to trace optically thin
free-free emission.  Power-law least squares fits to the flux densities of
the east and west inner lobes indicate spectral indices of $-0.17\pm0.08$
and $-0.21\pm0.08$, respectively. The best-fit spectral indexes for the
east and west outer lobes are $-0.22\pm0.03$ and $-0.32\pm0.03$,
respectively.  In each case the error bars correspond to 1-$\sigma$
uncertainty derived from the method described in \citet{Lampton1976ApJ},
and are based on flux uncertainties that do not include calibration errors.
We note that the values of the spectral indices are not far from $-0.1$,
which is the value expected for optically thin thermal free-free emission.
Furthermore, spectral indices between 5 and 9 GHz are even flatter (closer
to $-0.1$) compared with those calculated including the 17 and 19 GHz
fluxes, suggesting that a fraction of the flux at the high frequencies is
being resolved out by the interferometer.  We conclude that the emission
from the lobes does not have spectral indices significatively lower than
$-0.1$, as it would be expected in the case of optically thin synchrotron
radiation. Further interferometer observations with a more complete
uv-plane coverage will allow us to better recover the extended emission
associated with the lobes and to determine whether or not some fraction of
it can be attributed to non-thermal processes.

\subsubsection{Sources 1, 2 , and 3}

We detect source 1 at 9, 17, and 19 GHz. Because of the absence of IR or 3
mm counterparts, it is possible that its
nature is extragalactic, possibly a GHz peaked radio-galaxy, as observed in other regions of star formation \citep{Rodriguez2014AJ,Dzib2015ApJ}. According to
the model of extragalactic radio-source counts of \citet[at 15.7
  GHz]{AMI2011MNRAS.415.2708}, the probability of detecting one or more
radio sources brighter than 200 $\mu$Jy inside the primary beam of the ATCA
at 17 GHz ($140$\arcsec\ FWHM) is $\approx0.2$.

Source 3 exhibits a peak flux density at 17 GHz  of $1.2$ mJy. Since the
probability of finding one or more radio sources brighter than $1$ mJy
inside the 17 GHz primary beam is $\lesssim0.05$, it is unlikely that source
3 has an extragalactic origin.  A more likely possibility is that source
3 corresponds to free-free emission from a low-mass YSO \citep[see, for
  example,][]{AMI2012MNRAS}, presumably forming part of the young embedded
cluster associated with IRAS 16562$-$3959. This free-free emission is
characterized by a rising spectrum until the turnover frequency, which in
this case might be close to $20$ GHz.  If this were the case, we would
expect a 0.85 mJy source at 3 mm. Unfortunately, at 3 mm source 3 apparently blends
 with the $\sim2$ mJy source 9 from \citet{Guzman2014ApJ}
located $1\farcs2$ (less than one beam size) to the east.

Finally, source 2 is associated with the IR object GLIMPSE
G345.4977+01.4668 \citep{Benjamin2003PASP}, the 3 mm source 18 in
\citet{Guzman2014ApJ}, and the 6 GHz methanol maser MMB345.498+1.467
\citep{Caswell2010MNRAS}. Our observations provide new support for this
source being a HC \hii\ region associated with a young high-mass
star ($L_{\rm bol}\sim10^4\Lsun$). Keeping recombination equilibrium of
this HC \hii\ region requires an ionizing photon flux $\ge10^{45}$
s$^{-1}$, consistent with a B2 star \citep{Thompson1984ApJ}. The spectral
index measured between 17 and 99 GHz is 1.0, similar to that observed at 3
mm \citep{Guzman2014ApJ}.

%\clearpage
\subsection{Jet and Shock Parameters}
%R -> 5.2 Rsol, M -> 15 Msol, Vesc=1050 kms

%{\subsubsection{Shock Parameters and Jet Density}\label{sec-shock}}

The five radio sources \Oe, \Ie, \Iw, W$_1$, and \Ow\ are thought to trace
shocks associated with the protostellar jet from G345.49+1.47. However, it
is not clear whether these shocks are generated within the jet or in the
interaction zones with the ambient clump or with previously ejected
 material filling the outflow cavities.
%% In the following, we
%% assume that the outer lobes are tracing shocks associated with the
%% interaction between jet material and the molecular clump. Based only on its
%% location, we propose that W$_1$ correspond to an inner working surface.
%% The nature of the inner lobes is in part more uncertain because they
%% display slightly smaller PMs compared with the OW lobe.  However, the
%% results from \mbox{S.\ J.\ D.\ Purser} et al.\ (in preparation) compared with our observations
%% at 17 GHz indicate that the velocity in the plane of their sky of the inner
%% lobes is $\sim500$ \kms.
%
%% The molecular clump IRAS 16S562$-$3959 that hosts G345.49+1.47 was modeled
%% by \citet{Guzman2010ApJ}.  According to this model, the density and
%% temperature of the clump IRAS 16562$-$3959 at a distance $r$ from
%% G345.49+1.47 is given by
%% %X0
%% $\rho_c(r)=2.5\times10^{5}(\text{0.1 pc}/r)^{1.9}\mu$ cm$^{-3}$ and
%% $T_c(r)=50(0.1/r)^{0.4}$ K, respectively. Here,  $\mu=1.3{\rm m}_H$ is the 
%% average atomic mass in grams  assuming solar abundance.
%% %
%% The mean density at the distances of 
%% % assuming 45deg de incli.
%% the outer and inner lobes are 
%% $\rho_{c,{\rm out}}=6.56\times10^4\mu/\left(\sin(i)\right)^{1.9}$ and
%% $\rho_{c,{\rm inn}}=3.13\times10^6\mu/\left(\sin(i)\right)^{1.9}$ cm$^{-3}$, 
%% respectively, where $i$ is the inclination of the jet
%% respect to the line of sight.  
%% These values are likely
%% upper limits to the ambient density  at
%% the location of the lobes because these are possibly tracing shocks
%% occurring inside the outflow cavities. 
Theoretical models
show that the density inside outflow cavities could be between 100--1000 times
lower than the average density of the clump
\citep[e.g.,][]{Zhang2013ApJ}.

We  model the jet as having a cylindrical cross-section of area
$A_j=\pi R_j^2$, density $\rho_j$, and velocity $V_j$.  This highly
supersonic ($\mathcal{M}_j\sim100$) jet is ramming through an
stationary ambient medium of density $\rho_a$ and driving a detached
bow-shock traveling through the ambient medium at velocity $V_s$. The
shock speed of the detached shock is not necessarily $V_s$ --- except
at its apex --- because a large fraction of the shock front 
corresponds to weaker, oblique shocks. As a first approximation, we 
assume that the bow-shock sweeps a transverse area $A_s$. Conservation
of momentum implies that \citep{Chernin1994ApJ}
\begin{equation}
(V_j-V_s)^2\rho_jA_j=V_s^2\rho_a A_s~~.\label{eq-ram}
\end{equation}
On the other hand, the escape velocity from a 15 \Msun\ star --- the
estimated mass of the HMYSO --- of radius 5.2 $R_\odot$
\citep[][]{Mottram2011ApJL,Davies2011MNRAS} is $V_{\rm
  esc}\approx1050$ \kms.  Under a wide range jet acceleration
mechanisms, we expect the velocity of the jet not exceeding (at least
by a large factor) the escape velocity of the accreting compact object
\citep{Ferreira2006AA}. In fact, most jets appearing in several
astrophysical situations have velocities comparable with $V_{\rm esc}$
\citep{Livio2009pjc}.  In the case of G345.49+1.47, $V_{\rm esc}$ is
within a factor of $\sim2$ compared with the projected velocities
measured toward the \Oe\ and \Ow\ lobes, $390$ and $520$ \kms,
respectively.  Hereafter, we use the average and the unbiased standard
deviation between these two values, $460\pm90$ \kms, as the projected
shock velocity and its uncertainty, respectively.  We can express this
projected shock velocity as $V_s\sin(i)$, where $i$ is the inclination
angle between the jet direction and the line of sight. In
\citet{Guzman2011ApJ} we estimated $i=80\arcdeg$ from the molecular
outflow. However, we prefer to leave $i$ as a free parameter because
this estimation was obtained using relatively low angular resolution
data and an extremely simple geometrical model of the outflow. Note
also that $i$ may vary along the jet due
to precession, which  may help to 
explain the  difference of transverse velocities between the inner and
the outer lobes.  We summarize the conditions expected between the
shock, jet, and escape velocities by
\begin{equation} 
460\pm90~\kms/\sin(i)\approx V_s\le V_j\lesssim V_{\rm esc}\approx1050~\kms~~.%
\label{eq-vvv}
\end{equation}
These conditions  imply that $i\ge26\arcdeg$ and, in
combination with Equation \eqref{eq-ram},
\begin{equation}
\frac{\rho_a}{\rho_j}=\alpha\left(\frac{V_j}{V_s}-1\right)^2\le(1.6\pm0.7)\alpha~~,\label{eq-roj}
\end{equation}
where $\alpha=A_j/A_s$.  The mass
and momentum rates produced by one side of the jet, on the other hand,
are given by
\begin{align}
\dot{M}_j&=\pi R^2_j\rho_jV_j~~,\label{eq-Mjdot}\\
\dot{P}_j&=\pi R_j^2\rho_jV_j^2=\dot{M}_jV_j~~,\label{eq-Pjdot}
\end{align}
respectively. 

We define $\epsilon$ as the fraction of the jet's momentum being  shed to the ambient medium through the shock. We  approximate this
fraction using \citep{Chernin1994ApJ}
\begin{equation}
\epsilon=\frac{\rho_aV_s^2A_s}{\rho_jV_j^2A_j}=%
\frac{1}{\alpha}\left(1-\frac{V_s}{V_j}\right)^2~~,\label{eq-f}
\end{equation} 
where the second equality is derived from Equation \eqref{eq-ram}.  We
can constrain the rate of momentum being injected in the ambient
medium ($\epsilon\dot{P}_j$) using the radio emission of the lobes
($F_\nu$) under the assumption that it corresponds to shocked ambient
material using
\begin{equation}
\left(\frac{F_\nu d^2}{\rm Jy~kpc^2}\right) = {2.5 \epsilon\dot{P}_j}%
\left(\frac{\nu}{\rm GHz}\right)^{-0.1}%
\left(\frac{V_s}{500~\kms}\right)^{-0.32}%
\left(\frac{T_e}{\rm 10^4 K}\right)^{0.45}{\rm\Msun~\kms~yr^{-1}}~~.%
\label{eq-pdotRad}
\end{equation}
This equation assumes that the radio emission corresponds to optically
thin and completely ionized free-free gas (see \citealp{Cabrit1992AA}
and \citealp{Curiel1989ApLC}).  For the OW bow-shock (the OE shock is
similar), $F_{\rm 9 GHz}=5.7\pm0.6$ mJy. We assume that the
uncertainty is dominated by absolute calibration uncertainties, which
we conservatively assume to be 10\% (the measurement uncertainty is
much less).  Assuming $d=1.7$ kpc and $T_e=7000$ K
\citep{Guzman2014ApJ}, and using $V_s=460\pm90/\sin(i)$ \kms, we
conclude that
\mbox{$\epsilon\dot{P}_j=(9\pm1)\times10^{-3}(\sin(i))^{-0.32}$}
\Msun\ \kms\ yr$^{-1}$, where the uncertainty is derived from the
the mean shock velocity and  the radio flux at 9 GHz.

Further, we estimate the jet's dynamical time, $t_\text{dyn, j}$, as
the projected mean distance of the outer lobes from the HMYSO
($0.21\pm0.05$ pc) divided by its projected mean velocity of the outer
lobes, obtaining $t_\text{dyn, j}\approx450\pm140$ yr. The inclination
is not relevant for this calculation.  We then estimate how much
momentum the OW bow-shock has injected in the ambient medium as
$t_\text{dyn, j}\times\epsilon\dot{P}_j$, obtaining $4\pm1
(\sin(i))^{-0.32}$ \Msun\ \kms.  The momentum of the molecular outflow
associated with G345.49+1.47, on the other hand, is $2P_{\rm CO}\approx 15\pm2$ 
\Msun\ \kms \citep{Guzman2011ApJ}, where the
factor of two takes into account that we are summing the blue- and
red-shifted sides. We assume that the uncertainty of the outflow
momentum is dominated by a 10\% uncertainty in the absolute
calibration, but we stress that there are other sources of bias 
need to be kept in mind \citep{Downes2007AA,Cabrit1990ApJ}.  We follow
the prescriptions given by \citet{Downes2007AA} and drop the usual
$\cos(i)^{-1}$ inclination correction factor. This factor strongly
overestimates the real momentum of jet-driven molecular outflows
because an important fraction of the gas moves in transverse
directions with respect to the jet.

Because the momentum injected by the OW bow-shock is comparable within
a factor of 2 with that of one side of the molecular outflow, we
conclude that at most a few bow-shocks similar to the OW lobe could
have driven the entire molecular flow reported in
\citet{Guzman2011ApJ}.  In particular, since the sizes of the jet and
of the molecular outflow are similar, it is possible that the latter
was driven by the OE and OW shocks. We note, however, that the
dynamical time of the molecular outflow (calculated according to the
``perpendicular'' timescale defined in \citealp{Downes2007AA}) is
$\sim2000$ yr, larger than that of the outer lobes.
%
%% (15^2-9^2)**.5*1700AU/(15.3/1.92 kms)/2/3/yr
% 
In addition, the well developed bipolar near-IR cavities and
the K$_S$ emission extending much farther than the molecular flows
\citep[Fig.\ 5]{Guzman2011ApJ} seem to indicate that the ejection of
material is older than $t_\text{dyn, j}$. 
Further observations will determine whether or not the observed
extended K$_S$ emission corresponds to shocked H$_2$ evidencing 
older outflow activity.

The density of the pre-shocked material can be estimated from the peak
brightness of the lobes as follows.  Using the averaged 17-19 GHz data
we find that the peak intensity of the \Ow\ lobe is 
%$6.8\pm0.3\times10^{-4}$
$6.8\pm0.7\times10^{-4}$ Jy beam$^{-1}$ at a mean frequency of 18 GHz.
This intensity is equivalent to a brightness temperature of
$7.7\pm0.8$ K, using a beam solid angle of 0.38 arcsec$^2$.  At 9 GHz,
on the other hand, the measured peak brightness temperature of the OW
lobe is
%$17.3\pm0.2$ K 
$17\pm2$ K with a beam of 1.61 arcsec$^2$.  We can estimate the
intrinsic peak brightness temperature ($T_0$) of the OW lobe using the
relation $T_a=T_0/(1+\Omega_b/\Omega_s)$ \citep[chapter
  7]{Wilson2013ToR}, where $\Omega_s$ is the solid angle subtended by
the source, $\Omega_b$ is the beam size, and $T_a$ is the measured
brightness temperature.  Assuming that the emission comes from
optically thin free-free gas, the quotient between the measured peak
temperatures at frequencies $\nu_1$ and $\nu_2$ is given by
$(\nu_1/\nu_2)^{-2.1}(1+\Omega_{b,2}/\Omega_s)/(1+\Omega_{b,1}/\Omega_s)$.
From this relation and the peak temperature quotient between 9 and 18
GHz we derive 
%$\Omega_s\sim0.97\pm0.09~\text{arcsec}^2$. 
$\Omega_s\sim1.0\pm0.3~\text{arcsec}^2$. 
Hence,
%$T_0\approx10.7\pm0.5$ K.  
$T_0\approx11\pm3$ K. 
%Note that for this calculation we can use
%the measurement uncertainties 
Based on the free-free emission model
from shocks of \citet{Curiel1989ApLC}, the brightness temperature at
18 GHz is given in the optically thin limit by
\begin{equation}
T_b=0.105~{\rm K}\left(\frac{\rho_a}{10^3\mu~{\rm cm^{-3}}}\right)%
\left(\frac{V_s}{100~\kms}\right)^{1.68}\left(\frac{T_e}{\rm  10^4~K}\right)^{0.45}~~,%
\label{eq-cur}
\end{equation}
where $T_e$ is the temperature of the shock-ionized gas and $\rho_a$
is the ambient density.  Using $V_s=460\pm90/\sin(i)$ \kms,
$T_b=T_0=11\pm3$ K, and assuming $T_e=7000$ K we derive
$\rho_a=(9\pm4)\times10^{3}\mu\left(\sin(i)\right)^{1.68}$
cm$^{-3}$. This is also comparable with the density derived from
\citet[Figure 7]{Ghavamian1998ApJ}. Note that in this calculation we
are assuming that the shock velocity is given by the proper motion of
the OW lobe. This approximation is valid because we are using the peak
brightness temperature occurring at the apex of the bow-shock where
the shock front is close to normal.

Using this value of the pre-shocked material 
we obtain the effective size of the 
OW shock, $R_s$, defined as $\sqrt{A_s/\pi}$. Clearly, $\alpha=R_j^2/R_s^2$.
From Equations  \eqref{eq-Pjdot} and \eqref{eq-f} we derive
\begin{equation}
R_s=\left(\frac{\epsilon\dot{P}_j}{\pi\rho_aV_s^2}\right)^{1/2}\approx1400\pm200~\text{AU}~~,\label{eq-rs}
\end{equation}
where using the values of $\epsilon\dot{P}_j$ and $\rho_a$ estimated
above cancels the dependence on inclination and on $V_s$.  The value
for $R_s$ obtained in Equation \eqref{eq-rs} is within a 20\% of the
half angular size of the OW lobe, which is $\approx1\arcsec$,
equivalent to 1700 AU.  This consistency is not entirely trivial since
Equation \eqref{eq-rs} was not derived using measured sizes from the
images but using the proper motions, the total flux of the lobe, the
peak brightness temperature, and the quotient between the brightness
temperatures at two frequencies.  
Hereafter, we assume that $R_s$ ranges between 1400 and 1700 AU.
Using these values for $R_s$, we can
constrain $\alpha$ by assuming that the W$_1$ lobe corresponds to an
inner shock inside the jet. Since W$_1$ is unresolved and our smallest
synthesized beam has a FWHM with an equivalent physical size of $960$
AU, we assume that the transverse size of the jet ($2R_j$) is smaller
than half of the beam size, or $R_j\le240$ AU.

We can derive a rough lower limit for $R_j$ by noting that the
spectrum of the emission from the W$_1$ knot does not seem to arise
from optically thick gas.  While the flux density of W$_1$ --- fixed
by the observations, see Table \ref{tab-flux} --- is $\propto n_e^2R_j^3$, 
the optical depth of the knot  is given by $\tau\propto n_e^2 R_j$.
By requiring $\tau\le1$, we derive that $R_j\ge50\pm10$ AU.
Combining the previously derived constrains on $R_s$ and $R_j$, we finally derive
$9\times10^{-4}\le\alpha\le0.03$.

Finally, under the circumstances suggested in this section,
the brightness of the bow shocks and its associated Mach disk should be
similar \citep{Hartigan1989ApJ}. We can estimate the distance between the
Mach disk and the bow shock using the expression $\xi R_jc^0_s/2V_j$
\citep{Raga1998ajop}, where $c^0_s\sim10$ \kms\ is the sound speed of the
ionized gas and $\xi$ is close to unity for $\rho_j\approx\rho_a$. Since
$c^0_s/V_j\le0.02\sin(i)$ for G345.49+1.47, we expect the Mach disk and
bow shock emission to be blended in our observations. Based on this, we rule
out W$_1$ as being the Mach disk associated with the \Ow\ lobe.

\subsection{Mass Accretion Rate}

We can obtain a lower bound on  the mass outflow rate $\dot{M}_j$ of  the jet 
using Equations \eqref{eq-vvv},
\eqref{eq-roj}, \eqref{eq-Mjdot}, \eqref{eq-cur}, and \eqref{eq-rs}
%\begin{equation}
%\dot{M}_j=\frac{\epsilon\dot{P}_j}{\epsilon V_j}=\frac{\epsilon\dot{P}_j}{\alpha^{-1}\left(1-V_s/V_j\right)^2V_j}
%\ge2.7\times10^{-5}\alpha\left(\sin(i)\right)^{-0.32} \Msun~\text{yr}^{-1}~~.
%\end{equation} 
\begin{equation}
\begin{aligned} 
\dot{M}_j=\pi R_j^2\rho_jV_j\ge&\frac{\pi R_s^2\rho_aV_s}{1.6\pm0.7}\\
=&(1.2\pm0.8)\times10^{-5}\sin(i)^{0.68}\Msun~\text{yr}^{-1}~~.
\end{aligned}\label{eq-MjdotEst}
\end{equation}
To estimate the accretion rate onto the HMYSO, $\dot{M}_{\rm acc}$, we can
assume it is related with the jet's mass outflow rate through
$2\dot{M}_j=f_a\dot{M}_{\rm acc}$, where and $f_a$ typically ranges between
0.1--0.4
\citep[e.g.,][]{Tomisaka1998ApJ,Banerjee2006ApJ,Seifried2012MNRAS}.
Unfortunately, with the current data it does not seem possible to give an
upper limit on $\dot{M}_j$ because the density of the jet is not
constrained by the observations. In principle, it could be possible that
$\rho_j\gg\rho_a$ (with a large $\dot{M}_j$) and $\epsilon\ll1$, that is, a very dense jet piercing
the clump depositing a very small fraction of its momentum in the medium.

We can constrain from below the accumulated mass in the central object using
\begin{equation}
M_\star=2 \int_0^{t_\star} \frac{\dot{M}_j}{f_a}dt\ge
\frac{12\pm8}{f_a}\left(\frac{t_\star}{\rm Myr}\right)\sin(i)^{0.68}\Msun~~,\label{eq-maccum}
\end{equation}
where $t_\star$ is the HMYSO age. The last inequality in Equation
\eqref{eq-maccum} uses \eqref{eq-MjdotEst} and assumes that
$\dot{M}_j$ and $f_a$ have been constant during the entire HMYSO
lifetime. Since G345.49+1.47 is in the HC \hii\ region phase, we
assume its age is $\approx10^5$ yr \citep{Guzman2012ApJ}.  Although
with  quite 
large uncertainties, we can conclude that the jet characteristics
are roughly consistent with disk accretion rates of the order of a few
times $10^{-4}$ \Msun\ yr$^{-1}$ (assuming $f_a=0.1\text{--}0.2$),
which are sufficient to account for the current mass of the HMYSO.

{\section{SUMMARY}\label{sec-sum}}

We have observed the HMYSO G345.49+1.47 in radio continuum at 5, 9, 17, and
19 GHz using the ATCA between October 2014 and May 2015.  Our main
results are summarized as follows:
\begin{enumerate}
\item By comparing observations separated by 6 years, we determine that the
  PMs of the \Oe, \Ie, \Iw, and \Ow\ lobes are $48\pm13$, $35\pm10$,
  $17\pm11$, and $64\pm12$ mas yr$^{-1}$, respectively. Assuming a distance
  of 1.7 kpc, these PMs correspond to velocities of $390\pm100$, $280\pm80$, $140\pm90$, and $520\pm100$ \kms\ in the plane of the sky, respectively.
\item The PMs of the lobes are directed away from G345.49+1.47.  In
  addition, the morphology of the outer lobes --- especially of that of the
  western outer lobe --- are consistent with detached bow shocks.
\item We interpret these results as evidence that the radio lobes are
  produced in working surfaces associated with a highly collimated
  protostellar jet.  The radio lobes match the IR extended emission which
  is likely tracing the illuminated inner outflow cavity containing the jet.
  The jet's velocity is comparable with the HMYSO escape velocity, which is
  also the case for jets detected toward low-mass YSOs.  We propose that a
  mechanism of jet ejection and disk accretion similar to that of low-mass
  stars is acting in G345.49+1.47.  The presence of ionizing radiation or
  the HC \hii\ region does not hinder this process.
\item We observe emission arising from a previously undetected
  ionized knot (W$_1$) in the jet path located between the outer 
  and inner west lobes. 
\item We determined that the size spectral index of the HC \hii\ region
  (source \C) is $-1.1^{+0.5}_{-0.4}$,
  compatible with a bipolar (and possibly biconical) ionized
  region. 
\item The momentum injected by the bow-shocks in the past $\sim500$ yr is
  comparable with the observed momentum of the molecular outflow. We
  propose that the previously reported bipolar molecular outflow reflect
  only a relatively recent fraction of the history of ejection from
  G345.49+1.47 \citep{Guzman2011ApJ}. Further observations will determine
  whether or not there are indications of older ejection of material
  located farther from the HMYSO.   
\item {\textrm{The characteristics of the jet are roughly consistent with disk
  accretion rates of the order of $10^{-4}$ \Msun\ yr$^{-1}$, which are
  sufficient to account for the current estimated mass of G345.49+1.47 in
  $\sim10^5$ yr.}}
\end{enumerate}

\acknowledgements A.E.G.\ acknowledges support from FONDECYT through
grant 3150570. G.G.\ acknowledges support from CONICYT through project
PFB-06. The authors thank an anonymous referee for the constructive
comments and suggestions which helped to improve this manuscript.

\appendix

{\section{Angular Resolution Matching}\label{sec-match}}

In order to compare the features in maps made with different beam shapes,
it is convenient to transform the images to a common beam.  This is
important for disentangling real morphological differences from artifacts
introduced by the instrument and the observing technique.  When both images
are characterized by Gaussian and radially symmetric beams, the beam sizes
are matched by convolving the image associated with the smaller beam,
characterized by a $\text{FWHM}=\theta$, with a symmetric Gaussian of
$\text{FWHM}=\delta$, such that $\theta^2+\delta^2=\Theta^2$, where
$\Theta$ is the FWHM of the larger Gaussian beam. The larger Gaussian beam
is  the optimal common beam shape which can be attained without 
deconvolving the images.

Finding an optimal convolving  beam when the two images are characterized
by elliptical Gaussian beams is more involved. This is usually the case in
interferometry, where the images' resolution are determined by their
synthesized beams.  In order to reach a common resolution, one possibility
is to convolve the images with each other's beams. However, this solution
is not optimal and it does not match with the procedure described above for
circular beams.

In this appendix we derive a solution which is optimal in the sense 
it minimizes the sum in quadrature of the major and minor semiaxes of the
convolving beams. Denoting by $\alpha$, $B_{\rm maj}$, and $B_{\rm min}$
the P.A., major, and minor FWHM axes, respectively, of an elliptical
Gaussian beam, we define 
$\textbf{M}_1:=R(\alpha_1) \left(\begin{smallmatrix}B^2_{1,\rm maj} & 0\\ 0 & B^2_{1,\rm min}
  \\ \end{smallmatrix}\right) R(-\alpha_1)$ 
where $R(\alpha_1)$ is the rotation matrix associated with $\alpha_1$.
Subscripts 1 and 2 refer to quantities for each one of the images.  Let
$\textbf{D}=\textbf{M}_1-\textbf{M}_2$ and $d_1$ and $d_2$ its eigenvalues.
We define $\textbf{B}_1=R(\alpha_D)\Lambda_1 R(-\alpha_D)$, where
$R(\alpha_D)$ is the rotation matrix associated with the normalized
eigenvectors of $\textbf{D}$ and $\Lambda_1$ is a diagonal matrix defined
as $(\Lambda_1)_{ii}=-\min{(d_i,0)}$ for $i=1,2$. We define $\textbf{B}_2$
in the same way, but with diagonal terms equal to $\max{(d_i,0)}$. Matrices
$\textbf{B}_1$ and $\textbf{B}_2$ define the Gaussians that match the beams
of both images via convolution with $\textbf{M}_1$ and $\textbf{M}_2$,
respectively (that is,
$\textbf{B}_1+\textbf{M}_1=\textbf{B}_2+\textbf{M}_2$).  We still need to
prove that they are the optimal solutions.

Clearly, $\textbf{B}_1$ and $\textbf{B}_2$ are positive semi-definite (PSD)
matrices and $\textbf{B}_2-\textbf{B}_1=\textbf{D}$.  Every PSD matrix
represents a (maybe degenerate) Gaussian beam.  We now prove that among all
other PSD matrices ${X}$ such that ${X}+\textbf{D}$ is PSD, $\textbf{B}_1$ minimizes the
sum in quadrature of its eigenvalues.  Let $r_j$ be the eigenvectors
associated with the negative (or negatives) eigenvalues of $\textbf{D}$ and
$F({X})$ the sum of the squares of the eigenvalues of a matrix ${X}$. Then
$F({X})\ge\sum_{j}(r_j^T{X}r_j)^2$. Since $\textbf{D}+{X}$ is PSD, we have that
$d_j+r_j^T{X}r_j\ge0$, and because $d_j<0\le r_j^T{X}r_j$, we conclude that
$(r_j^T{X}r_j)^2\ge d_j^2$. Therefore $F({X})\ge\sum_jd_j^2=F(\textbf{B}_1)$.
Similarly, for each PSD matrix $Y$ such that $Y-D$ is PSD, $\textbf{B}_2$
minimizes $F$.  Therefore, if ${X}$ and ${Y}$ are PSD and fulfill
$\textbf{M}_1+{X}=\textbf{M}_2+{Y}$, then $F({X})+F({Y}) \ge F(\textbf{B}_1)+F(\textbf{B}_2)$.  Therefore, $\textbf{B}_1$ and
$\textbf{B}_2$ minimize the sum in quadrature of their eigenvalues and
fulfill $\textbf{B}_1+\textbf{M}_1=\textbf{B}_2+\textbf{M}_2$ 
\citep[see also][\S5.2]{horn2012matrix}.

Matrices $\textbf{B}_1$ and $\textbf{B}_2$ define the convolving beams.
The P.A. is $\alpha_D$ and the FWHM axes squared are given in the diagonal
terms of $\Lambda_i$.  If $\textbf{B}_1$ (respectively, $\textbf{B}_2$) is
$0$, then we only need to convolve the second (first) image by the beam
described by the parameters of $\textbf{B}_2$ (${\textbf{B}}_1$).  If one
of the convolving beam's axes is zero (as it may happen when both beams are
of similar size but have different orientations), then the convolving beam
is degenerate and infinitely narrow in one direction. In practice, we limit
the minimum axis of the convolving beam to 3 pixels.

%%%%%%%%%%%%%%%%%%%%%%%%%%%%%%%%%%%%%%%%%%%%%%%%%%%%%%%%%%%%%%%%%%%%%%
%% Bibliography
%%%%%%%%%%%%%%%%%%%%%%%%%%%%%%%%%%%%%%%%%%%%%%%%%%%%%%%%%%%%%%%%%%%%%%
\bibliographystyle{apj}
%\bibliography{bibliografia} %% comentar esta linea antes de submitir
\bibliography{manuscript} %% descomentar esta linea antes de submitir

\end{document}